\documentclass[twocolumn]{aastex63}

\usepackage{amsmath}

\usepackage{times}

\usepackage{graphicx}

\usepackage{tikz}

\usepackage[T1]{fontenc}
\usepackage{aecompl}
\usepackage{calligra}

\usepackage{comment}

\DeclareMathAlphabet{\mathcalligra}{T1}{calligra}{m}{n}
\DeclareFontShape{T1}{calligra}{m}{n}{<->s*[2.2]callig15}{}

\definecolor{colordisk}{rgb}{0.78,0.918,0.898}
\definecolor{color3}{rgb}{0.749,0.506,0.1765}

\usepackage{hyperref}
\hypersetup{
    pdfnewwindow=true,      
    colorlinks=true,       
    linkcolor=color3,      
    citecolor=color3,        
    filecolor=magenta,      
    urlcolor=color3         
}

\def\apjl{ApJL}               
\def\apjs{ApJS}               
\def\aap{A\&A}                

\defcitealias{DL:2016}{DL16}

\usepackage{color}
\usepackage{xcolor}

\newcommand{\OK}[1]{\textcolor{red}{[OK]}}

\def\wb{\omega_\mathrm{b}}
\def\qd{q_\mathrm{d}}

\def\eb{e_\mathrm{b}}
\def\rcav{r_\mathrm{cav}}

\def\qb{q_\mathrm{b}}
\def\ab{a_\mathrm{b}}
\def\Mb{M_\mathrm{b}}
\def\qb{q_\mathrm{b}}
\def\ob{\Omega_\mathrm{b}}
\def\rin{r_\mathrm{in}}

\def\rs{r_\mathrm{s}}

\def\rout{r_\mathrm{out}}
\def\css{c^2_\mathrm{s}}

\def\wp{\omega_P}
\def\wq{\omega_Q}
\def\ss{\zeta}

\def\bomb{\overline{\omega}_\mathrm{b}}

\usepackage{xifthen}
\definecolor{red1}{rgb}{0.7, 0.15, 0.15}
\newcommand*{\needcite}[1]{
    \ifthenelse{\equal{#1}{}}{
        {\color{red1}[?]}
    }{
        {\color{red1}[#1]}
    }
}

\usepackage{lineno}

\usepackage[makeroom]{cancel}

\usepackage{comment}

\begin{document}

\shorttitle{Forced CBD Eccentricity}
\shortauthors{Grcić, D'Orazio, Pessah}

\title{Insights from Analytical Theory of Eccentric Circumbinary Disks II. \\ Forced Modes and Resonance for Precessing Binaries}

\author[0000-0001-5301-2564]{Marcela Grcić}
\affiliation{Niels Bohr International Academy, Niels Bohr Institute, Blegdamsvej 17, DK-2100 Copenhagen Ø, Denmark}
\email{marcela.grcic@nbi.ku.dk}

\author[0000-0002-1271-6247]{Daniel J. D'Orazio}
\affiliation{Space Telescope Science Institute, 3700 San Martin Drive, Baltimore, MD 21218, USA}
\affiliation{Department of Physics and Astronomy, Johns Hopkins University,
3400 North Charles Street, Baltimore, Maryland 21218, USA}
\email{dorazio@stsci.edu}
\affiliation{Niels Bohr International Academy, Niels Bohr Institute, Blegdamsvej 17, DK-2100 Copenhagen Ø, Denmark}

\author[0000-0001-8716-3563]{Martin E. Pessah}
\affiliation{Niels Bohr International Academy, Niels Bohr Institute, Blegdamsvej 17, DK-2100 Copenhagen Ø, Denmark}
\email{mpessah@nbi.ku.dk}

\begin{abstract}
An eccentric, unequal-mass binary induces forced eccentricity in a circumbinary disk through the non-axisymmetric component of its gravitational potential. Building on the theory of free (i.e., unforced) eccentric modes, we develop a semi-analytical framework to describe this response in two-dimensional, locally isothermal disks with a power-law surface density profile.
We show that the disk eccentricity is governed by the competition between pressure and the binary quadrupole potential, leading to two distinct regimes. In quadrupole-dominated disks, the eccentricity oscillates about the forced eccentricity of a test particle, $E\sim r^{-1}$, with an amplitude and wavelength set by the disk aspect ratio. In pressure-dominated disks, the eccentricity departs qualitatively from the test-particle limit and follows a universal radial scaling $E\sim r^{-2}$, consistent with recent numerical results.
Resonant amplification occurs when the binary forcing frequency matches the eigenfrequency of a free eccentric disk mode. In the limit of a non-precessing binary, this reduces to the previously identified zero-frequency resonance, for which we derive an analytic criterion and map its dependence on disk and binary parameters.
We extend the framework to massive disks by including the disk's gravitational potential and allowing binary apsidal precession. 
We conjecture that the cavity size, for eccentric, non-equal-mass binaries, can be set such that the ground free eccentric mode of the disk has an eigenfrequency equal to the binary precession frequency. In other words, the disk cavity adjusts until the lowest-order trapped eccentric mode resonates with the forcing from the precessing binary.
\end{abstract}

\section{Introduction}

Binaries often form and evolve in gas-rich environments, which can result in the formation of circumbinary accretion disks \citep[CBDs; ][]{1986_Boss,2007_hayasaki,2023MNRAS.523.4353E}. 
These disks can influence the evolution of the binary through the exchange of energy and angular momentum, as well as through mass accretion
\citep[e.g.,][]{Armitage_Natarajan_2002,
Munoz_Lai_2019, 
Zrake_Tiede_MacF_Haiman_2021, 2023_Lai_Munoz_review, Siwek_orbevo+2023, Dittmann_Ryan_2024, KITP_CC:2024,Franchini_eccSG_2024, ClyburnZrake:2026}. 
Similarly, the evolution of the disk is affected by the binary \citep[e.g.,][]{Artymowicz_Lubow_1994, Rafikov_2013, Rafikov_2016, Dan-gaps, Ragusa+_2020, Noble+_2021, Siwek+_2023}.
Modeling the binary-disk interaction is crucial for predicting binary populations, from stellar systems to supermassive black hole binaries \citep[e.g.,][]{Izzard,Valli+2024,Siwek_PTAsorbevo+2024, MurrayDuffell:2025,2026_Unger}, and for deciphering electromagnetic signatures that can identify or characterize accreting binary systems
\citep[e.g.,][]{Farris_Duffel_2014,Tofflemire_DQTau+2017,2022_gutierrez, DOrazioCharisi:2023, DOrazio_binlite+2024, 2024_Cocchiararo, Tiwari_RMHD_I+2025}.

The CBD–binary interaction is complex and has been the subject of numerous analytical studies and (magneto-)hydrodynamical simulations. These show that the disk can develop significant eccentricity and reach a steady state in which it precesses uniformly, extending to radii of up to ten times the binary separation 
\citep[e.g.,][]{MacFadyen_Milosavljevic_2008, Shi+_2012, Dorazio_Haiman_2013, Dunhill+2015, Dan-transition, Miranda_Munoz_Lai_2017, ML2020, Noble+_2021, Siwek+_2023, KITP_CC:2024, Tiwari_RMHD_I+2025}. Furthermore, 2D hydrodynamical simulations have found that the disk precession and/or eccentricity can vanish for intermediate binary eccentricity values, and is strongly tied to changes in binary orbital evolution \citep{Miranda_Munoz_Lai_2017, Dan-transition},
a phenomenon which could be attributed to Lindblad resonances \citep{Miranda_Munoz_Lai_2017} or to the adopted inner boundary conditions \citep{Thun_plus_2017}, but is still not understood.

In spite of the complexity of the problem, important insights have been gained with semi-analytical approaches, relying on increasing degrees of sophistication.
Within the framework of perturbation theory, steady-state eccentric disks can be described as eccentric modes in the CBD, with the disk precession rate being the mode eigenvalue \citep{Ogilvie_2001, GO2006,2023_Lai_Munoz_review}. When only the axisymmetric component of the binary gravitational potential is considered, these modes -- usually referred to as ``free'' or ``unforced'' modes -- 
have been used to successfully describe some aspects of eccentric CBDs that arise in the fully-non-linear problem \citep[e.g.,][]{Shi+_2012, Miranda_Munoz_Lai_2017, ML2020, KITP_CC:2024}.
In \citet{p1}, we recently carried out a systematic study of free modes, establishing a coherent framework to better understand how their properties depend on the binary and disk parameters.

A natural extension of this framework is to include the non-axisymmetric $(m=1)$ component of the binary gravitational potential, which arises for an unequal-mass, eccentric binary. This component drives forced eccentricity in the disk, giving rise to so-called ``forced'' modes that precess at the same rate as the binary potential. The properties of these modes, and in particular the conditions under which their amplitudes are enhanced, remain less well understood.

We motivate the need for a deeper understanding of forced eccentric modes with three observations. First, in the absence of self-gravity and pressure, the forced eccentricity of a circumbinary test particle scales as $e_{\rm f} \propto r^{-1}$ \citep{Moriwaki_Nakagawa_2004,Lee_Paele_2006}. 
However, hydrodynamical simulations by \cite{Siwek+_2023} show that the eccentricity profile of a gaseous disk around an eccentric binary can differ substantially from this test-particle expectation. Second, \citet{Lubow2022} showed that the forced eccentricity of a disk can be resonantly amplified when a non-precessing eccentric binary excites a zero-frequency free eccentric mode. This suggests a broader interpretation: forced eccentricity should be amplified whenever the binary forcing frequency matches the eigenfrequency of a free eccentric disk mode. Third, if the disk is sufficiently massive, its gravitational potential can drive apsidal precession of the binary, shifting the forcing frequency away from zero and thereby modifying the resonance condition.

Given these considerations, our goal is to build on the framework developed in \citet{p1} to understand the structure and resonant amplification of forced eccentric modes in circumbinary disks. We first consider non-precessing eccentric binaries, for which the resonance condition reduces to the existence of a non-trivial zero-frequency free mode and can be characterized analytically. We then extend the framework to include binary apsidal precession driven by a massive disk\footnote{Our approach can also accommodate relativistic binary precession.}, allowing us to examine how disk mass shifts the location of resonances and modifies the forced eccentricity profile.

Our study of forced eccentric modes in CBDs is organized as follows. In Section \ref{sec:methods}, we set up the problem. We use the perturbative semi-analytic methods developed by \cite{Tremaine_2001} and \cite{GO2006} to describe the CBD steady-state evolution defined by the disk eccentricity distribution and its precession frequency. This approach allows us to isolate the effects of the disk pressure, and the different components of the gravitational potential. In Section \ref{sec: 4}, we consider disks around non-precessing eccentric binaries. We find the forced disk eccentricity and the conditions for resonant eccentricity evolution in the disk. Then, in Section \ref{sec: 5}, we relax the assumption of a massless disk by adding the corresponding disk gravitational potential component. We calculate the binary precession rate caused by the disk mass, and find the resulting CBD evolution. We conclude with a discussion of implications for CBD accretion studies and future directions in Section \ref{sec: Discussion and Conclusion}.

\section{Methods}\label{sec:methods}

We consider a binary and its surrounding gaseous disk with the following properties. The binary mass is $\Mb=M_1+M_2$, and the binary mass ratio is $\qb=M_2/M_1 \leq 1$, where $M_1$ and $M_2$ are the masses of the primary and the secondary. The binary eccentricity is $\eb$, and the binary semi-major axis is $\ab$. The orbital frequency of the binary is $\ob=\sqrt{G \Mb/\ab^3}$, where $G$ is the gravitational constant. The mid-plane of the disk is coplanar with the binary orbital plane. The disk scale height is $H(r)$. We assume a constant disk aspect ratio $h=H/r=\mathrm{const}$.

To calculate the precession frequency of the disk, and its eccentricity, we use the perturbative method formulated by \cite{Ogilvie_2001}. This approach relies on defining the disk eccentricity through perturbations with respect to a circular reference disk state. The reference state is characterized by no radial velocity ($u_\mathrm{ref}=0$), axial symmetry ($\partial_\phi X_\mathrm{ref}=0$, where $\phi$ is the azimuthal coordinate), and time independence ($\partial_t X_\mathrm{ref}=0$) of all fluid quantities $X_\mathrm{ref}$ (i.e., the pressure $p$, surface density $\Sigma$, and angular and radial velocities $v$ and $u$). The reference state and the perturbations with respect to it are then expanded into a series in the small parameter $\epsilon \propto h$. The perturbations take the form $X_\mathrm{pert}(r,\phi,t)=\Re \left[\Tilde{X}(r)e^{-i(\phi-\omega t)} \right]$, where $\Tilde{X}$ are Fourier coefficients corresponding to the azimuthal wavenumber $m=1$ \citep{Lee_Dempsey_Lithwick_2019}. The disk eccentricity function $E(r,t)$ is then defined through the velocity perturbations as $\Tilde{v} \approx r\Omega_0 E(r)/2$ (angular velocity) and $ \Tilde{u} \approx ir\Omega_0 E(r)$ (radial velocity). $\Omega_0=\sqrt{G\Mb/r^3}$ is the dominant term in the orbital frequency expansion in radius of a disk fluid element (see below Eq. (\ref{eq: potential binary m=0})). These definitions allow us to write the resulting eccentricity equation in terms of the eccentricity function $E(r)$ and reference values for the disk density and pressure.
 
The time evolution of the disk eccentricity is given by the eccentricity equation: 
\begin{equation}\label{eq: eccentricity equation in terms of f_p and f_g with forcing}
\begin{split}
    \Sigma r^2\Omega_0 \frac{\partial E}{\partial t}&= f_p(E,r)+f_{\mathrm{g},m=0}(E,r) +f_{\mathrm{g},m=1}(E,r) ,
\end{split}
\end{equation}
where pressure function $f_p(E,r)$ captures pressure effects, while functions $f_{\mathrm{g},m=0}(E,r)$ 
and $f_{\mathrm{g},m=1}(E,r)$ describe the $m=0$ (axisymmetric) and $m=1$ components of the binary gravitational potential, respectively. We neglect viscous effects, though bulk viscosity can be incorporated in the linear analysis as an imaginary part of the adiabatic coefficient \citep{GO2006}. 

We write the disk eccentricity as $E(r,t) =E(r) e^{i\omega t}$, where
$E(r)$ is the radial eccentricity profile and $\omega$ is the uniform precession rate. In this steady-precession regime, the time derivative in the first term of Eq. (\ref{eq: eccentricity equation in terms of f_p and f_g with forcing}) reduces to
\begin{equation}
\label{eq: steady state time derivative of E(r,t) }
\partial_t E(r,t) = i \omega E(r,t) = i \omega E(r) e^{i \omega t},
\end{equation}
so that each annulus of the disk precesses coherently at the same rate $\omega$.

In this work, we limit the discussion to 2D locally isothermal disks\footnote{We restrict our analysis to a 2D locally isothermal disk, as this is the standard assumption in the hydrodynamical simulations to which we compare our results.}, for which the pressure influences the eccentricity of the disk via \citep{TO2016}:
\begin{equation}\label{eq: pressure forcing term f_p for the eccentricty equation }
    \begin{split}
        f^\mathrm{2D,is}_p=& \frac{i}{r}\frac{\partial}{\partial r} \left(\frac{1}{2}\Sigma \css r^3 \frac{\partial E}{\partial r} \right) + \frac{ir}{2}\frac{d }{d r}\left( \Sigma \css\right) E\\
&-\frac{i}{2r}\frac{\partial}{\partial r}\left( \Sigma\frac{d \css}{dr}r^3 E\right).
    \end{split}
\end{equation}
The 2D approximation assumes that both the vertical velocity of the fluid and its perturbations are zero. The locally isothermal approximation is equivalent to setting a fixed radial temperature profile. With a constant disk aspect ratio $h$, the equivalent sound speed profile is given as $\css\propto 1/r$.
For 3D and/or adiabatic disks, the pressure effects differ from the one described by Eq. (\ref{eq: pressure forcing term f_p for the eccentricty equation }). The expressions for $f_p(E,r)$ for such disks can be found in \cite{TO2016}, see also \citet{p1}.

We approximate the $m=0$ component of the binary gravitational potential as:
\begin{equation}\label{eq: potential binary m=0}
    \begin{split}
        \Phi_{m=0}=&
        -\frac{G\Mb}{r} + \Phi_{Q}  \\
        =& -\frac{G\Mb}{r} -\frac{G\Mb \ab^2}{r^3}\frac{\qb}{4(1+\qb)^2} \left(1+\frac{3}{2}\eb^2 \right).
    \end{split}
\end{equation}
The first term in Eq. (\ref{eq: potential binary m=0}) sets the value of $\Omega_0=\sqrt{G\Mb/ r^{3}}$. The second term is the quadrupole component of the gravitational potential, the lowest-order non-Keplerian axisymmetric contribution of the gravitational potential $\Phi_{m=0}$, whose effect on the CBD evolution is given by
\begin{equation}\label{eq: gravitational forcing term for the eccentricty equation}
    f_{\mathrm{g},m=0}\equiv -\frac{i\Sigma }{2}\frac{\partial}{\partial r}\left[r^2 \frac{\partial \Phi_{m=0}}{\partial r}  \right]E.
\end{equation}
In some cases, in addition to the quadrupole binary potential, the gravitational potential from the disk mass is significant enough to be included in $\Phi_{m=0}$. We discuss this further in Section \ref{sec: 5}.

We approximate the non-axisymmetric $m=1$ component of the binary gravitational potential by
\begin{equation} \label{eq: m=1 pot}
\begin{split}
    \Phi_{m=1}=& \frac{G\Mb}{r}\frac{15}{16}\frac{\ab^3}{r^3} \times \\
    & \eb\left(1+\frac{3}{4}\eb^2\right)\qb\frac{(1-\qb)}{(1+\qb)^3} e^{i\omega_\mathrm{b}t},
    \end{split}
\end{equation}
where $\omega_\mathrm{b}$ is the binary precession rate. Its contribution to the eccentricity equation is
\begin{equation}\label{eq: gravitational forcing term w_bin}
    f_{\mathrm{g},m=1}\equiv \frac{i\Sigma }{2r}\frac{\partial}{\partial r}\left[r^2 \Phi_{m=1} \right].
\end{equation}
Because this is independent of $E$, it sets the scale for disk velocity perturbations to the reference state and so the amplitude of $E$.

Throughout this work, we assume a finite disk with a central cavity of size $\rin$ and an outer radius $\rout$. Following \citet{Ogilvie_2001} and \citet{2008Planetesimal}, we demand the Lagrangian pressure perturbations to vanish at the disk edges $\rin$ and $\rout$. For locally isothermal disks, this condition is equivalent to:
\begin{equation}\label{eq: the boundary condition - iso}
    \left. \left[ \frac{E}{(p/\Sigma)} \right]' \right|_{\rin}=\left. \left[ \frac{E}{(p/\Sigma)} \right]' \right|_{\rout}=0,
\end{equation}
where prime denotes a radial derivative. We assume a power-law radial density profile:
\begin{equation}\label{eq: sigma_zeta}
    \Sigma(r)=\Sigma_0 \left( \frac{r}{\ab} \right)^\zeta,  \rin \leq r \leq \rout,
\end{equation}
where $\zeta$ is the density-power-law exponent, and $\Sigma_0$ is a global constant. We assume that the density of a CBD decreases with increasing radius, so we set $\zeta<0$.

The exact form of the pressure-induced precession frequency, obtained by substituting $\Phi_{m=0}=\Phi_{m=1}=0$ into Eqs. (\ref{eq: eccentricity equation in terms of f_p and f_g with forcing})-(\ref{eq: gravitational forcing term w_bin}), depends on the form of the density profile $\Sigma(r)$. To use a measure of the pressure-induced precession frequency that is not restricted to a single density profile, we define \citep{p1}:
\begin{equation}\label{eq: pressu freq}
    \wp \equiv \frac{h^2}{2} \left(\frac{\rin}{\ab}\right)^{-3/2}\ob,
\end{equation}
as the measure of the pressure-induced precession frequency at the disk inner edge. Note that the definition Eq. (\ref{eq: pressu freq}) differs from \citet{ML2020} by a factor of $1/2$.

We also define 
\begin{equation}\label{eq: quadr freq}
    \wq\equiv \frac{3\qb}{4(1+\qb)^2} \left(1+\frac{3}{2}\eb^2 \right) \left(\frac{\rin}{\ab}\right)^{-7/2}  \ob,
\end{equation}
as the precession frequency at the disk inner edge caused by the quadrupole gravitational potential of the binary \citep{GO2006}. Equation (\ref{eq: quadr freq}) can be obtained by substituting $p=\Phi_{m=1}=0$ into Eqs. (\ref{eq: eccentricity equation in terms of f_p and f_g with forcing})-(\ref{eq: gravitational forcing term w_bin}).

\section{Non-Precessing Eccentric Binary}\label{sec: 4}
We initially assume that the disk mass is negligible compared to the mass of the binary, so that the binary neither precesses ($\overline{\omega}_\mathrm{b}=0$) nor evolves, and we thus neglect the gravitational potential of the disk.
For the disk eccentricity $E(r,t)$ to be a steady-state solution to the forced eccentricity equation, with the non-axisymmetric potential arising from the eccentric binary, the disk precession frequency has to be equal to that of the binary \citep{Lubow2022}. 
Hence, for the first case that we consider, of a non-precessing binary, we substitute $\omega=\overline{\omega}_\mathrm{b}=0$ into Eqs. (\ref{eq: eccentricity equation in terms of f_p and f_g with forcing})-(\ref{eq: gravitational forcing term w_bin}).

\subsection{Forced Test Particle -- Prelude}

In the limit of a pressure-less disk ($p=h=0$), the solution to the eccentricity equation given by Eqs. (\ref{eq: eccentricity equation in terms of f_p and f_g with forcing})-(\ref{eq: gravitational forcing term w_bin}) is equal to the forced eccentricity of a test particle moving in the gravitational potential of the binary $\Phi_{m=0}+\Phi_{m=1}$
\citep{Moriwaki_Nakagawa_2004, Bromley_Kenyon_2015,Lubow2022}:
\begin{equation}\label{eq: forced ecc}
    e_\mathrm{f}(r)=\frac{5}{4}\frac{1-\qb}{1+\qb}\frac{\ab}{r}\eb \frac{1+3/4\eb^2}{1+3/2\eb^2}.
\end{equation}
The forced eccentricity vanishes for a circular ($\eb=0$) or an equal-mass ($\qb=1$) binary. These are cases for which there is no non-axisymmetric term in the gravitational potential, i.e., $\Phi_{m=1}=0$. 

\subsection{Forced Gaseous Disk -- General Considerations}

If the pressure in the disk is not negligible, that is, if the disk has a finite thickness ($h>0$), the disk
eccentricity distribution $E(r)$ differs from one of a test particle given by Eq. (\ref{eq: forced ecc}). 
The solution to the forced eccentricity equation consists of a linear combination of the two homogeneous solutions $E_1(r)$ and $E_2(r)$ corresponding to $f_{g,m=1}=0$, plus a particular solution $E_3(r)$ that accounts for the non-axisymmetric forcing associated with $f_{g,m=1}\ne 0$ in Eq. (\ref{eq: gravitational forcing term w_bin}), i.e., 
\begin{equation}\label{full ecc sol}
    E(r)=\eta_\mathrm{1} E_1(r)+\eta_\mathrm{2} E_2(r)+ E_3(r).
\end{equation}
The constants $\eta_1$ and $\eta_2$ are fixed by applying the boundary condition in
Eq.(\ref{eq: the boundary condition - iso}) to the general solution $E(r)$ from Eq. (\ref{full ecc sol}), which gives
\begin{subequations}\label{betas_forced}
\begin{equation}\label{b1 forced}
    \eta_1=-\frac{\Psi_\mathrm{3,2}}{\Psi_\mathrm{1,2}},
\end{equation} 
\begin{equation}\label{b2 forced}
    \eta_2=\frac{\Psi_\mathrm{3,1}}{\Psi_{1,2}},
\end{equation}
\end{subequations}
where we define $\Psi_{i,j}$ as a measure of the value of the disk eccentricity and its derivatives at the disk edges:
\begin{equation}
\label{eq:zeta_12}
\begin{split}
    \Psi_{ij}\equiv& 
    \partial_r\left[E_{i} \frac{\Sigma}{p} \right]_{r_\mathrm{in}}
    \partial_r\left[E_{j}\frac{\Sigma}{p}\right]_{r_\mathrm{out}}\\ 
  &-\partial_r\left[E_{j}\frac{\Sigma}{p}\right]_{r_\mathrm{in}}
    \partial_r\left[E_{i}\frac{\Sigma}{p}\right]_{r_\mathrm{out}}
    \end{split},
\end{equation}
with $E_i$ and $E_j$ representing any of the three solutions $E_1$, $E_2$, or $E_3$ introduced in Eq. (\ref{full ecc sol}). For free modes (solutions to the homogeneous equation), $\Psi_{1,2}=0$ is simply the condition that needs to be satisfied for the system of equations with given boundary conditions to have a solution \citep[see][]{p1}.

By Eq. (\ref{full ecc sol})-(\ref{eq:zeta_12}), the eccentricity forcing provides a scale for the eccentricity absent in the case of free disk precession \citep{Ogilvie_2001,ML2020, p1}. To find the magnitude of the eccentricity scale, \cite{Lubow2022} studied adiabatic disks with a power-law-density profile $\Sigma \propto r^{-1/2}$ with a central cavity of size $\rcav=2\ab$, and for a binary mass ratio of $\qb=2/3$. \cite{Lubow2022} showed that the forced disk eccentricity takes resonant values for values of the disk aspect ratio $h$ for which $\omega=0$ is the precession frequency of a free/unforced disk, i.e., with $\Phi_{m=1}=0$.

In what follows, we expand the assumptions in \cite{Lubow2022} to consider locally isothermal disks, other density-power-law exponents $\zeta$, and different choices for the inner disk radius $\rin$ and binary mass ratio $\qb$. Within this larger context, we provide analytical results for sets of binary-disk parameters necessary to excite the extreme eccentricity values first discussed by \cite{Lubow2022}.

The homogeneous solutions $E_1(r)$ and $E_2(r)$ for eccentricity profiles in power-law-density disks around a non-precessing binary are given by \citep{p1}:
\begin{equation}\label{eq: non-precessing disks - full analytic solution for E(r)}
E_{1,2}= \left(\frac{\rs}{r}\right)^{\delta} J_{\pm\nu}\left(\frac{\rs}{r} \right)\, ,
\end{equation}
where the subscripts $+$ and $-$ refer to the solutions $1$ and $2$, respectively.
Here, the value of the Bessel parameter $\nu$ for 2D locally isothermal disks
\footnote{Adiabatic and 3D versions of Eqs. (\ref{eq: nu parameter})-(\ref{eq: r_s definition}) can be found in \cite{p1}.} is given by:
\begin{equation}\label{eq: nu parameter}
    \nu^2=\frac{\ss^2}{4}-\ss+1.
\end{equation}
The exponent $\delta$ is related to the power-law density exponent $\ss$ and, for locally isothermal disks, is:
\begin{equation}\label{eq: delta parameter}
    \delta = -\frac{\ss}{2}-1.
\end{equation}
The scale radius, $\rs$, is the radius at which quadrupole and pressure-induced precession frequencies would equal, i.e., $\omega_Q(r)=\omega_P(r)$.
The scale radius is proportional to the strength of the quadrupole gravitational potential and inversely proportional to the disk aspect ratio $h$:
\begin{equation}\label{eq: r_s definition}
    \rs=\sqrt{\frac{3 \qb }{2(1+\qb)^2}\left(1+\frac{3}{2}\eb^2 \right)\frac{1}{h^2}}\ab.
\end{equation}
In \cite{p1}, we showed that the quantity $\rs$ is related to the ratio of the quadrupole to pressure induced frequencies at the disk inner edge. Specifically, we substitute the definitions given by Eq. (\ref{eq: pressu freq}) and Eq. (\ref{eq: quadr freq}) into Eq. (\ref{eq: r_s definition}) to obtain:
\begin{equation}\label{eq: wp_eq}
    \left( \frac{\rs}{\rin}\right)^2=\frac{\wq}{\wp}.
\end{equation}

\subsection{A Concrete Example}

We illustrate the effect of binary eccentricity forcing on the disk eccentricity with an example. We adopt a model for which the analytical solutions for all the terms of the eccentricity solution in Eq. (\ref{full ecc sol}) are relatively simple: a locally isothermal 2D disk with a power-law-density exponent $\zeta=-1$. For such a disk, the homogeneous solutions are given by:
\begin{equation} \label{eq: example homogenous solution}
    \begin{split}
        E_1=& \left(\frac{\rs}{r}\right)^{1/2}J_{3/2}\left(\frac{\rs}{r} \right)\\
        E_2=& \left(\frac{\rs}{r}\right)^{1/2}J_{-3/2}\left(\frac{\rs}{r} \right)
    \end{split},
\end{equation}
and the particular solution is:
\begin{equation}\label{eq: example particular solution}
\begin{split}
    E_3(r)=&h^2 \frac{5}{3}\frac{(1-\qb)(1+\qb)}{\qb}\frac{\eb\left(1+\frac{3}{4}\eb^2 \right)}{\left(1+\frac{3}{2}\eb^2 \right)^2}\frac{r}{\ab}\\
    &+\frac{5}{4}\frac{1-\qb}{1+\qb}\eb \frac{1+3/4\eb^2}{1+3/2\eb^2}\frac{\ab}{r},\\
    =& e_\mathrm{f}(r)\left[ 1+2 \left(\frac{r}{\rs}\right)^2\right].
    \end{split}
\end{equation}
For an infinitely thin disk $h\rightarrow0$, the radial scale becomes infinite $\rs\rightarrow\infty$, see Eq. (\ref{eq: r_s definition}), and the above particular solution reduces to the forced eccentricity of a test particle given by Eq. (\ref{eq: forced ecc}), $E_3(r)\rightarrow e_\mathrm{f}(r)$, as expected.

In characterizing the properties of the solutions for the disk eccentricity, we first distinguish between two different limits in terms of the magnitude of the quadrupole-to-pressure ratio at the disk inner edge given by Eq. (\ref{eq: wp_eq}). We define quadrupole-dominated systems as those for which the quadrupole-induced precession frequency at the disk inner edge is higher than the pressure-induced one; $\rs/\rin\gg 1$. In the quadrupole-dominated limit, we approximate the solutions given by Eq. (\ref{eq: example homogenous solution}) as:
\begin{subequations} \label{eq: example approx E hom thin disk}
    \begin{equation}
        E_1(r)\approx  -\sqrt{\frac{2}{\pi}}\cos{\left(\frac{\rs}{r}\right)}  , \quad \rs/r \gg 1
    \end{equation}
    \begin{equation}
        E_2(r) \approx  -\sqrt{\frac{2}{\pi}}\sin{\left(\frac{\rs}{r}\right)} , \quad \rs/r \gg 1.
    \end{equation}
\end{subequations}

We define pressure-dominated systems as those for which the pressure-induced precession frequency at the disk inner edge is higher than the quadrupole-induced one; $\rs/\rin\ll 1$. In the pressure-dominated limit, we approximate the solution given by Eq. (\ref{eq: example homogenous solution}) as:
\begin{subequations}\label{eq: example approx E hom thick disk}
    \begin{equation}
        E_1(r) \approx \frac{1}{3}\sqrt{\frac{2}{\pi}}\left(\frac{\rs}{r}\right)^2 , \quad \rs/r \ll 1,
    \end{equation}
    \begin{equation}
    \begin{split}
        E_2(r) \approx & -\sqrt{\frac{2}{\pi}}\Bigg[ \left(\frac{r}{\rs}\right)+\frac{1}{2}\left(\frac{\rs}{r}\right)\\
        & -\frac{1}{8}\left(\frac{\rs}{r}\right)^3\Bigg] , \quad \rs/r \ll 1,
        \end{split}
    \end{equation}
\end{subequations}
where we have kept terms up to $\mathcal{O}(r^{-3})$ in radius.

 We next explore the two limits separately to highlight the differences in the resulting eccentricity profile for the two regimes, using the above described example system.

\subsection{Quadrupole-Dominated Disks}\label{subsec: quad dominated}

We assume a disk for which the quadrupole gravitational potential dominates over pressure at the inner edge, i.e., $\rs/\rin\gg 1$.
For near-circular, near-equal-mass binaries, and an inner radius of a few binary separations $\ab$, this condition is satisfied for disks with aspect ratio $h\ll 0.1$. Later, in \S \ref{subsection: Resonant Disk Eccentricity}, we discuss a more precise division of quadrupole and pressure-dominated limits.

To write the total forced eccentricity solution given by Eq. (\ref{full ecc sol}) using Eqs. (\ref{eq: example homogenous solution})-(\ref{eq: example particular solution}), we require constants $\eta_1$ and $\eta_2$. To find $\eta_1$ and $\eta_2$, we use the boundary condition given by Eq. (\ref{eq: the boundary condition - iso}) and the approximations for the homogeneous solutions in Eq. (\ref{eq: example homogenous solution}) at the inner and the outer disk radii.

At the disk inner edge, per our quadrupole dominated limit assumption, we can approximate the homogeneous solution with Eq. (\ref{eq: example approx E hom thin disk}). At the disk outer radius, which we assume is large enough for the approximation $\rs/\rout\ll 1$ to be valid, we approximate values of homogeneous solution with Eq. (\ref{eq: example approx E hom thick disk}).

Using these assumptions, we substitute Eq. (\ref{eq: example particular solution}) and Eqs. (\ref{eq: example approx E hom thin disk})-(\ref{eq: example approx E hom thick disk}) into Eqs. (\ref{betas_forced})-(\ref{eq:zeta_12}) to find the values of the two constants $\eta_1$ and $\eta_2$:
\begin{subequations}\label{etas example}
    \begin{equation}
    \begin{split}
        \eta_1 \approx & \sqrt{\frac{\pi}{2}}\frac{5}{2}\frac{1-\qb}{1+\qb}\eb \frac{1+3/4\eb^2}{1+3/2\eb^2}\frac{\ab}{\rs}\cot{\left(\rs/\rin\right)} \\
        =&  \sqrt{2\pi} e_\mathrm{f}(\rin)\frac{\rin}{\rs} \cot{\left(\rs/\rin\right)},
        \end{split}
    \end{equation}
    \begin{equation}
    \begin{split}
        \eta_2\approx& \sqrt{\frac{\pi}{2}}\frac{5}{2}\frac{1-\qb}{1+\qb}\eb \frac{1+3/4\eb^2}{1+3/2\eb^2}\frac{\ab}{\rs}\\
        =&  \sqrt{2\pi} e_\mathrm{f}(\rin)\frac{\rin}{\rs} .
        \end{split}
    \end{equation}
\end{subequations}
We finally substitute Eqs. (\ref{eq: example particular solution})-(\ref{eq: example approx E hom thin disk}) and Eq. (\ref{etas example}) into Eq. (\ref{full ecc sol}) to obtain the approximate expression for the forced eccentricity of the disk:
\begin{equation}\label{eq: forced ecc tot, grav lim}
\begin{split}
    E(r)\approx& \frac{5}{2}\frac{1-\qb}{1+\qb}\eb \frac{1+3/4\eb^2}{1+3/2\eb^2}\Bigg[ -\frac{\ab}{\rs}\cot{(\rs/\rin)}\cos{\left(\frac{\rs}{r}\right)} \\
    &-\frac{\ab}{\rs} \sin{\left(\frac{\rs}{r}\right)}  +\frac{\ab r}{\rs^2}+\frac{1}{2}\frac{\ab}{r}    \Bigg]; r/\rs< 1\\
    =& e_\mathrm{f}(r)\Bigg[ -2\frac{r}{\rs}\cot{(\rs/\rin)}\cos{\left(\frac{\rs}{r}\right)} \\
    &-2\frac{r}{\rs} \sin{\left(\frac{\rs}{r}\right)}  +\frac{2 r^2}{\rs^2}+1    \Bigg] ; r/\rs< 1.
\end{split}
\end{equation}

The expression for the forced eccentricity given by Eq. (\ref{eq: forced ecc tot, grav lim}) is valid in the inner parts of the disk, for  $\rs/r\gtrsim 1$, which we will show is valid up to $100\ab$ for the examples below. We consider the $\rs/r\lesssim 1$ limit in \S \ref{subsection: Pressure-dominated systems} and first focus on the highly eccentric part of the disk, given by Eq. (\ref{eq: forced ecc tot, grav lim})

\begin{figure}[h]
    \centering
    \includegraphics[scale=0.95]{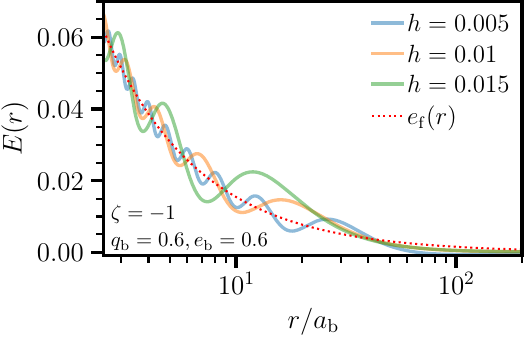}
    \caption{Disk eccentricity profiles $E(r)$ for an eccentric binary ($\eb=0.6$) with a binary mass ratio of $\qb=0.6$, and for three values of the disk aspect ratio: $h=0.005$ (solid blue line), $h=0.01$ (solid orange line), and $h=0.015$ (solid green line). We plot the forced test particle eccentricity given by Eq. (\ref{eq: forced ecc}) (dotted red line).  The disk is a 2D locally isothermal disk with a density profile $\Sigma=\Sigma_0 (r/\ab)^{-1}$, and inner and outer radii $\rin=2.5\ab$ and $\rout=200\ab$. For these, quadrupole-dominated systems, the forced disk eccentricity $E(r)$ oscillates around the eccentricity of a test particle $e_\mathrm{f}(r)$. The magnitude of the oscillations increases with increasing value of the disk aspect ratio $h$.}
    \label{fig: E(r) forced omega=0 h variation}
\end{figure}

In Figure \ref{fig: E(r) forced omega=0 h variation}, we plot the forced eccentricities for a disk around a non-precessing, eccentric binary ($\eb=0.6$) with a binary mass ratio $\qb=0.6$. The forced eccentricity of a test particle $e_\mathrm{f}(r)$ given by Eq. (\ref{eq: forced ecc}) is shown with a dotted red line. We also plot with solid lines the solutions for the forced eccentricity profiles for three different values of the disk aspect ratio ($h=0.005,0.01,0.015$), obtained by numerically solving the eccentricity equation, i.e., Eqs. (\ref{eq: eccentricity equation in terms of f_p and f_g with forcing})-(\ref{eq: the boundary condition - iso}). In the inner disk region, the forced eccentricity profiles for a disk with pressure ($h\neq 0$) oscillate around the forced particle ($h=0$) eccentricity given by Eq. (\ref{eq: forced ecc}), as predicted by Eq. (\ref{eq: forced ecc tot, grav lim}). These oscillations are given by $\cos{(\rs/r)}$ and $\sin{(\rs/r)}$ and, by Eq. (\ref{etas example}), their amplitudes are proportional to $\rin/\rs$. Since, by  Eq. (\ref{eq: r_s definition}), $\rin/\rs \propto h$, the amplitude of radial oscillations seen in Figure \ref{fig: E(r) forced omega=0 h variation} decreases for thinner disks. Increasing the pressure in the disk causes larger deviations of the disk eccentricity from $e_\mathrm{f}(r)$. Similarly, the wavelength of these oscillations is proportional to $1/\rs$, and thus increases with increasing disk aspect ratio $h$. The forced eccentricity of a test particle, $e_\mathrm{f}(r)$, and so $E(r)$, decreases with $\eb$; in the limit of $\eb\rightarrow0$ or $\qb \rightarrow 1$, the eccentricity profile vanishes everywhere, $E(r)\rightarrow 0$, unless $\omega=0$ is the eigenfrequency of a free mode \footnote{These free modes are discussed in detail in \cite{p1}.}.  Next, we discuss the implications of non-precessing free modes on the forced eccentricity.

\subsubsection{Resonant Disk Eccentricity}\label{subsection: Resonant Disk Eccentricity}

The solutions shown in Figure (\ref{fig: E(r) forced omega=0 h variation}) above constitute a relatively small deviation from the particle forced eccentricity. As shown by \cite{Lubow2022}, the disk eccentricity can reach much larger values in systems for which the binary and disk parameters allow resonant values of the disk eccentricity to be excited. In what follows, we elucidate the origin of this resonant behavior.

The origin of the resonant behavior can be understood from the structure of the boundary-value problem. In the absence of forcing, non-trivial eccentric disk solutions exist only when the homogeneous problem admits a solution that satisfies the boundary conditions, corresponding to $\Psi_{1,2}(\omega)=0$, which identifies the eigenfrequencies of the free eccentric modes studied in \citet{p1}. In the presence of forcing, the eccentricity can be written as the sum of a particular solution and a linear combination of the homogeneous solutions, with coefficients that scale as $\Psi_{3,i}/\Psi_{1,2}$, see Eqs. (\ref{full ecc sol})-(\ref{betas_forced}). As a result, the forced response is strongly amplified when $\Psi_{1,2}(\omega)\rightarrow 0$.
Physically, this corresponds to the case in which forcing frequency matches the eigenfrequency of a free eccentric disk mode. For a non-precessing binary, then, the amplitude of the forced mode diverges if a non-trivial non-precessing free mode exists \citep{Lubow2022}. With this understanding, we use our analytical solutions for the disk eccentricity to show explicitly how and where such resonances arise. Importantly, the resonance condition can be expressed in terms of the ratio $\rs/\rin = \sqrt{\omega_Q/\omega_P}$.

The resonant values of the disk eccentricity for a power-law density disk occur when $\Psi_{1,2}=0$, which, for power-law-density disks, occurs when \citep{p1}:
\begin{equation}
\label{eq: zeta approximation non-precessing power law disks}
    \Psi_{1,2} \approx \frac{\rs}{\rin}-\frac{\nu \pi}{2}-\left(n+\frac{1}{4} \right)\pi=0,
\end{equation}
where $n\geq 0$ is the order of the non-precessing free mode, equal to the number of local extrema of $E_1(r)$, and where $\nu$ is defined by Eq. (\ref{eq: nu parameter}). 
The resonant values of the forced eccentricity are possible only for quadrupole-dominated systems because Eq. (\ref{eq: zeta approximation non-precessing power law disks}) can only be satisfied for $\rs/\rin \geq \pi (1+2\nu)/4$.
By Eq. (\ref{eq: zeta approximation non-precessing power law disks}), there is a largest value of disk aspect ratio, $h_0=h(n=0)$, for which a resonant eccentricity value is possible.

\begin{figure}[h!]
    \centering
    \includegraphics[scale=1]{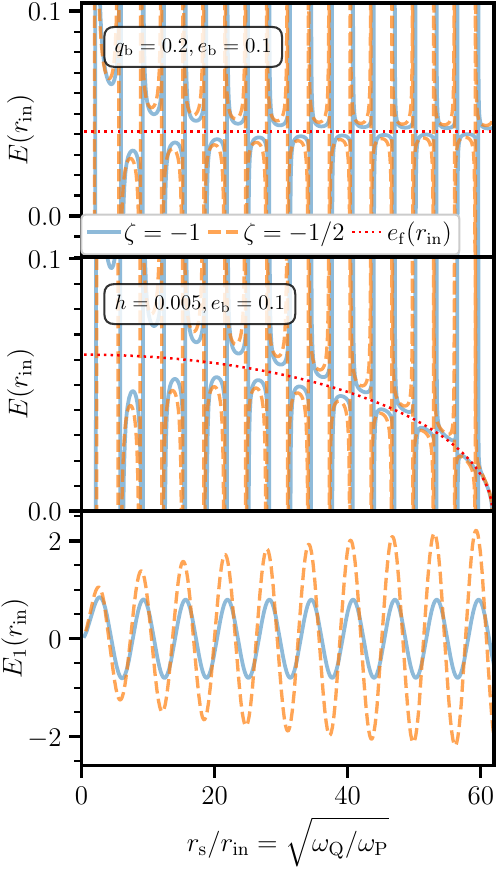}
    \caption{The value of the disk eccentricity at the disk inner radius $E(\rin)$ (upper panel) and the homogeneous solutions $E_1(r)$ (bottom panel) for quadrupole-to-pressure ratio values $0\leq \rs/\rin \leq 60$. We plot solutions for locally isothermal disks for the power-law-density exponent values $\zeta=-1$ (solid blue line) and $\zeta=-1/2$ (dashed orange line). We plot the values of the forced eccentricity of a test particle given by Eq. (\ref{eq: forced ecc}) (dotted red line). The forced disk eccentricity diverges for quadrupole-to-pressure values for which $E_1(r)$ takes extreme values. For non-resonant values, the disk eccentricity (upper and middle panels) converges to the value for the free-particle forced eccentricity (dotted red line) with increasing $\rs/\rin$ (decreasing disk aspect ratio in upper panel and increasing binary mass ratio in middle panel).
    The disk is a 2D locally isothermal disk with inner and outer radii $\rin=2\ab$ and $\rout=200\ab$. The values of the binary mass ratio and eccentricity are $\qb=0.2$ (upper panel) and $\eb=0.1$.}
    \label{fig: E(rin)-rs/rin forced precessing}
\end{figure}

In the top and middle panels of Figure \ref{fig: E(rin)-rs/rin forced precessing}, we plot the value of the disk eccentricity at the disk inner radius $E(\rin)$ against the values of $\rs/\rin$ for two values of the power-law-density exponent $\zeta=-1/2,-1$. In the upper panel, we plot the solutions for system parameter values $(\qb,\eb,\rin)=(0.2,0.1,2\ab)$, against the values of $\rs/\rin$ for disk aspect ratio values $0.004\leq h \leq 0.5$, corresponding to the quadrupole-to-pressure ratio of $0.5\lesssim \rs/\rin \lesssim 60$. In the middle panel, we set $(h,\eb,\rin)=(0.005,0.1,2\ab)$, and change the binary mass ratio value in range $10^{-5}\leq \qb \leq 0.99$, corresponding to the quadrupole-to-pressure ratio of $0.5\lesssim \rs/\rin \lesssim 60$. The lower end of $\qb$ values is included for the sake of obtaining results for $\rs/\rin$ values as low as in the top panel. But, the very low $\qb$ values are not consistent with the assumption of a gap of size $2\ab$ opening \citep{Dan-gaps}. In both panels, the value of $\zeta$ does not change the solutions much. The values of $\rs/\rin$ for which resonant disk eccentricity is realized are equidistant, as predicted by Eq. (\ref{eq: zeta approximation non-precessing power law disks}): $\left(\rs/\rin \right)_\mathrm{res}=\left(n+1\right)\pi$, where we have used $\nu=3/2$ for $\zeta=-1/2$.  In both panels, the leftmost resonance corresponds to the $n=0$ mode of the free non-precessing mode, while eccentricity spikes for progressively higher values of $r_s/\rin$ represent higher order modes, up to $n=18$ depicted here. 

In the top and middle panels of Figure \ref{fig: E(rin)-rs/rin forced precessing}, the dotted red line represents the value of the forced test particle eccentricity at $\rin$, given by Eq. (\ref{eq: forced ecc}). For increasing values of $\rs/\rin$, the values of the non-resonant forced eccentricity approach the values of the forced eccentricity of a test particle. This is in agreement with Figure \ref{fig: E(r) forced omega=0 h variation}. In the upper panel, $r_s/\rin$ is varied at fixed binary parameters, so $e_\mathrm{f}(\rin)$ is a constant. In the middle panel, $r_s/\rin$ is varied by changing the binary mass ratio, thus causing the value of $e_\mathrm{f}(\rin)$ to decreases with increasing binary mass ratio, approaching zero for $\qb=1$.

The condition for the existence of free non-precessing modes, i.e., $\Psi_{1,2}=0$, and resonance, can be approximated by Eq. (\ref{eq: zeta approximation non-precessing power law disks}). This approximate condition is functionally equal to finding local maxima of $E_1(\rs/\rin)$. We demonstrate this by plotting the homogeneous solution $E_1(r)$ given by Eq. (\ref{eq: non-precessing disks - full analytic solution for E(r)}) in the bottom panel of Figure \ref{fig: E(rin)-rs/rin forced precessing}. It is clear that the values of $\rs/\rin$ for which $E_1(\rin)$ has local extreme values, are the same values of $\rs/\rin$ for which the forced eccentricity diverges in the upper and middle panels.
The choice of the value for $\zeta$ has a strong effect on the magnitude of the unforced eccentricity, but not the resonant values of $\rs/\rin$, especially for high values of $n$. This is because, for high values of $n$, the resonance condition (Eq. (\ref{eq: zeta approximation non-precessing power law disks})) can be approximated as $\rs/\rin\approx n\pi$, which is not a function of $\zeta$.

\begin{figure}[!t]
\vspace{0pt}
$
\begin{array}{c}
\includegraphics[scale=0.99]{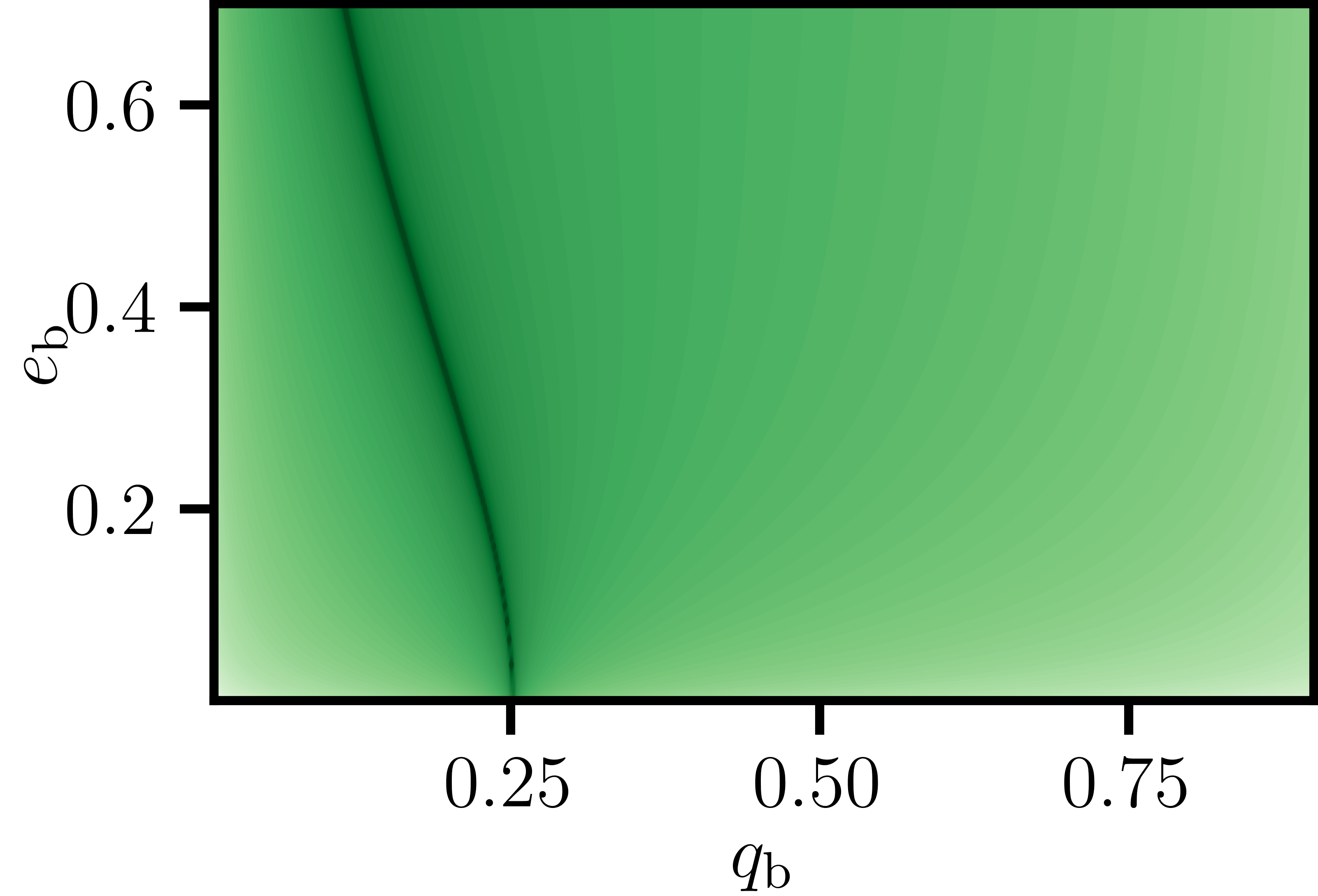} \\
\includegraphics[scale=0.99]{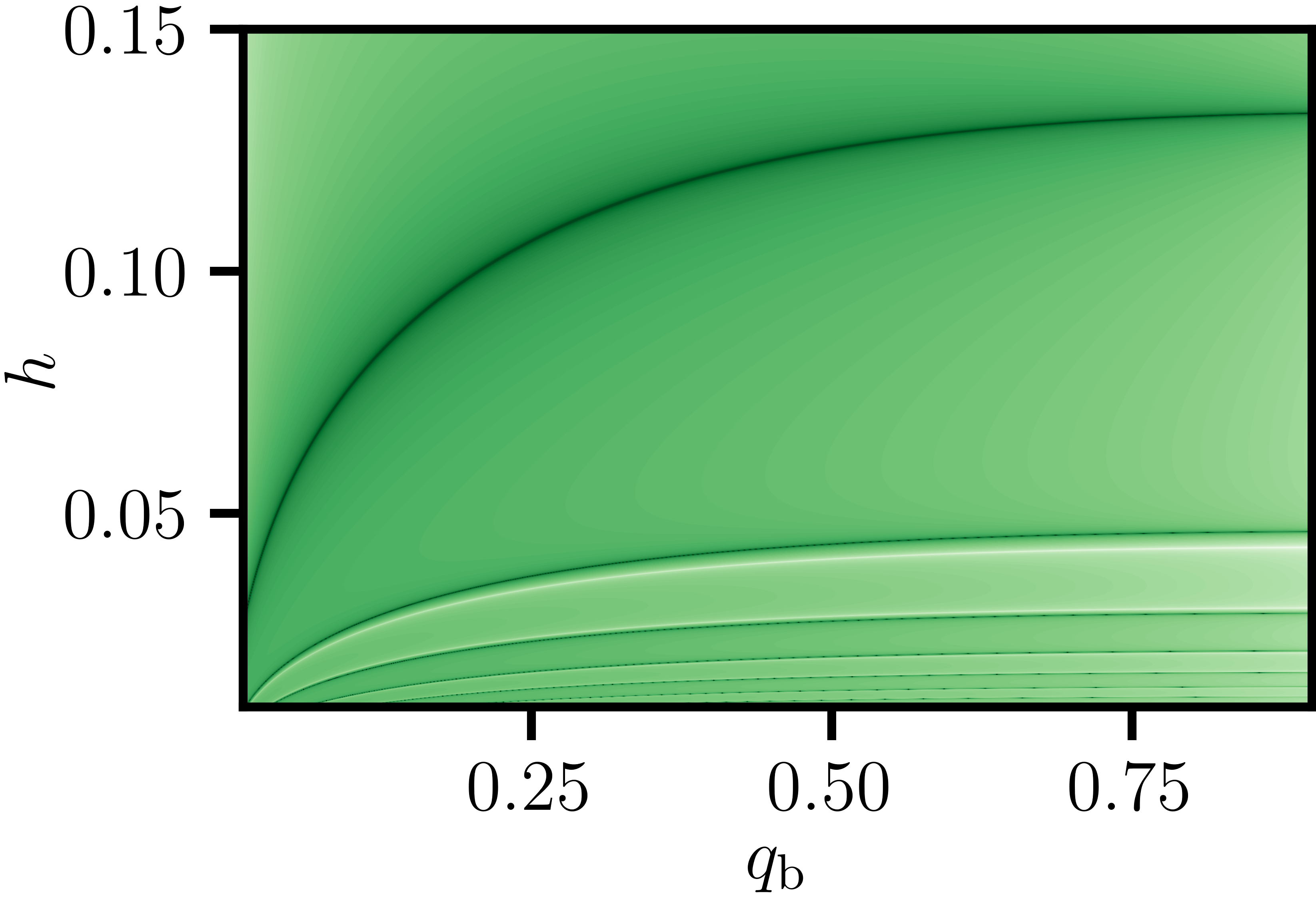}  \\
\includegraphics[scale=0.99]{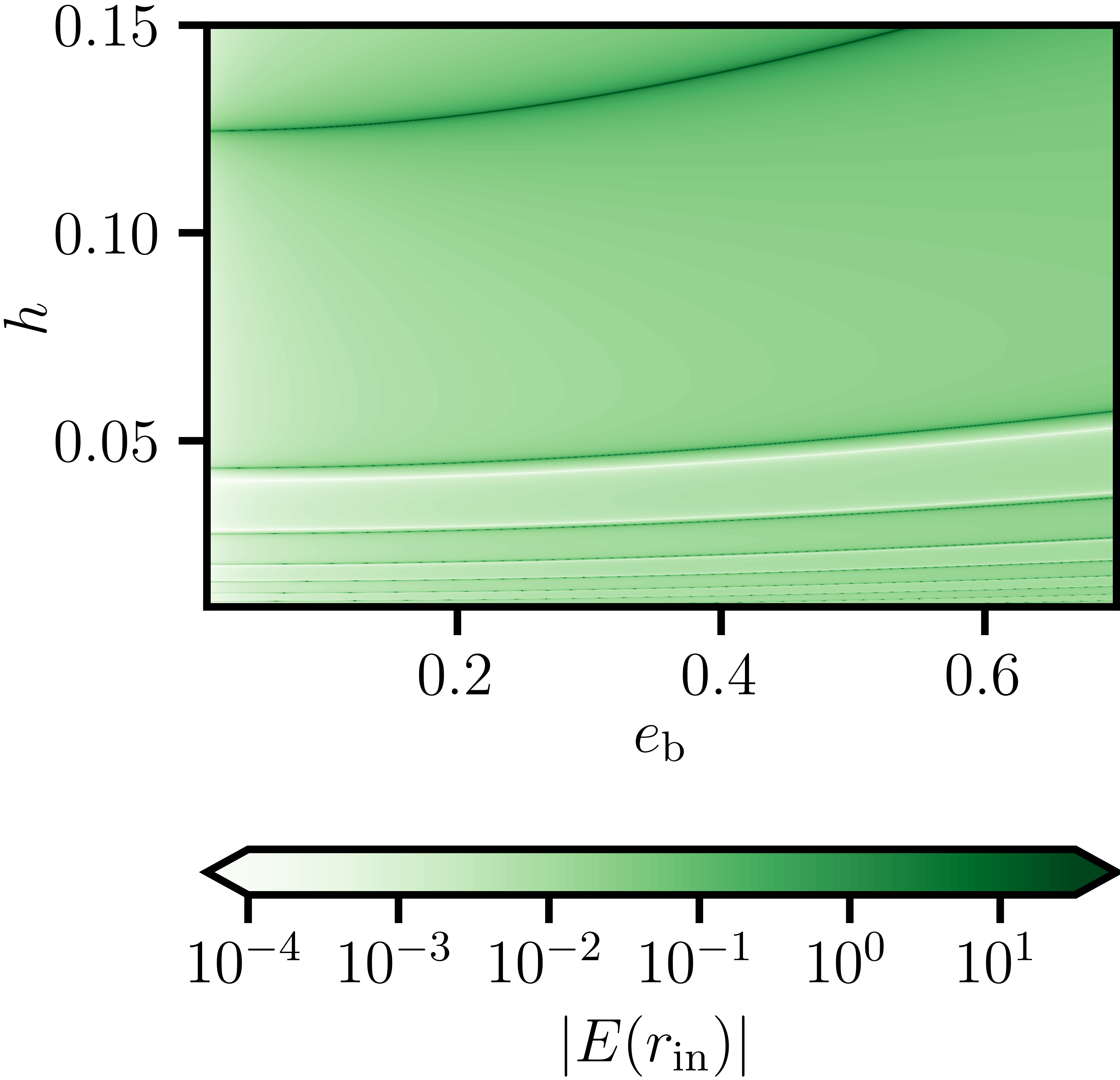}\\ 
\end{array}$
\caption{Values of the forced disk eccentricity at the disk inner edge $E(\rin)$ for varying values of the binary mass ratio and eccentricity with $h=0.1$ (panel 1); the binary mass ratio and the disk aspect ratio with $\eb=0.3$ (panel 2); the binary eccentricity and the disk aspect ratio with $\qb=0.9$ (panel 3). Dark green areas denote resonant parameters. The disk is a 2D locally isothermal disk with a density profile $\Sigma=\Sigma_0 (r/\ab)^{-1/2}$, and inner and outer radii $\rin=2.5\ab$ and $\rout=200\ab$. 
}
\label{fig:contour plot 1}
\end{figure}

There are four parameters that determine the solution to the eccentricity equation; $\eb$, $\qb$, $h$, and $\rin$. In Figure \ref{fig:contour plot 1}, we plot contours of disk eccentricity in the space spanned by binary eccentricity, the binary mass ratio, and the disk aspect ratio. Resonant values of the eccentricity are thin continuous dark green lines. Different dark green lines in the same panel correspond to different values of the free, non-precessing mode order $n$. Resonant eccentricities for lower values of the disk aspect ratio correspond to higher order modes. For higher order modes, resonant contours are thinner. This is because the range of parameters for which the disk is highly eccentric is much more narrow for higher order modes (see the top panel in Figure \ref{fig: E(rin)-rs/rin forced precessing}). Similarly, \cite{Lubow2022} shows that the width of resonant peaks in $h$ space increases with increasing value of $h$. The thin white lines denote parameters for which $E(\rin)=0$, seen in Figure \ref{fig: E(rin)-rs/rin forced precessing}.

In Figure \ref{fig:contour plot 1} (middle and bottom panels), the resonant lines delineate the regions where the eccentricity profile is of a different mode order (See Figure 7 in \cite{p1}). The most prominent (thickest) resonant line represents the $n=0$ mode boundary, and thus delineates the pressure-dominated and quadrupole-dominated regimes.

\subsection{Pressure-dominated Disks}\label{subsection: Pressure-dominated systems}

The eccentricity of a free non-precessing mode is given by $E(r)=\beta_1 E_1(r)+\beta_2 E_2(r)$, where $\beta_1$ and $\beta_2$ are constants. Since usually $\lim_{r \to \infty}E_2=\infty$, for large outer disk radii, the boundary conditions set $\beta_2=0$, and $E(r) \approx E_1(r)$ \citep{p1}. Specifically, in the pressure-dominated regime, the eccentricity of a free non-precessing mode is given as $E(r) \approx E_1(r) \propto (r_s/r)^{\delta+\nu}$. For a 2D locally isothermal disk, $\nu=\zeta/2-1$ (Eq.~\ref{eq: nu parameter}) and $\delta=-\zeta/2-1$ (Eq.~\ref{eq: delta parameter}), implying $E_1\propto r^{-2}$ (Eq.~\ref{eq: example approx E hom thick disk}). This scaling holds for all $\zeta<0$. We next show in an example that a similar argument holds for forced disk eccentricity; 
because of the boundary conditions, leading terms in $E_2$ and $E_3$ cancel, leaving the 
forced eccentricity approximately equal to $E(r) \approx E_1(r) \propto r^{-2}$.

For pressure-dominated systems, the homogeneous solutions can be approximated by Eq. (\ref{eq: example approx E hom thick disk}) for all $\rin\leq r\leq \rout$. So
we can substitute Eq. (\ref{eq: example particular solution}) and Eq. (\ref{eq: example approx E hom thick disk}) into Eq. (\ref{eq:zeta_12}) to find the values of constants $\eta_1$ and $\eta_2$ (Eq. (\ref{betas_forced})):
\begin{subequations}\label{etas example thick disk}
    \begin{equation}\begin{split}
        \eta_1 \approx &-\frac{3}{4}\sqrt{\frac{\pi}{2}}\frac{5}{2}\frac{1-\qb}{1+\qb}\eb \frac{1+3/4\eb^2}{1+3/2\eb^2}\frac{\ab}{\rin}\\
        =& -\frac{3}{2}\sqrt{\frac{\pi}{2}}e_\mathrm{f}(\rin),
        \end{split}
    \end{equation}
    \begin{equation}
    \begin{split}
        \eta_2\approx & \sqrt{\frac{\pi}{2}}\frac{5}{2}\frac{1-\qb}{1+\qb}\eb \frac{1+3/4\eb^2}{1+3/2\eb^2}\frac{\ab}{\rs}\\
        =& \sqrt{2\pi}\frac{\rin}{\rs}e_\mathrm{f}(\rin).
        \end{split}
    \end{equation}
\end{subequations}
We substitute Eq. (\ref{eq: example particular solution}), Eq. (\ref{eq: example approx E hom thick disk}) and Eq. (\ref{etas example thick disk}) into Eq. (\ref{full ecc sol}) to find  the total forced disk eccentricity profile:
\begin{equation}\label{eq: forced approx example}
    \begin{split}
        E(r)= \frac{e_\mathrm{f}(\rin)}{2}\left[\frac{1}{2}\frac{\rs^2\rin}{r^3}-\frac{\rs^2}{r^2} \right].
    \end{split}
\end{equation}
The dominant term in the disk eccentricity profile given by Eq. (\ref{eq: forced approx example}) is proportional to $r^{-2}$. 
Terms from $E_2$ proportional to $r^{-1}$ and $r^1$ cancel equivalent terms in $E_3$ because of the imposed boundary conditions.

\begin{figure}[h]
    \centering
    \includegraphics[scale=0.95]{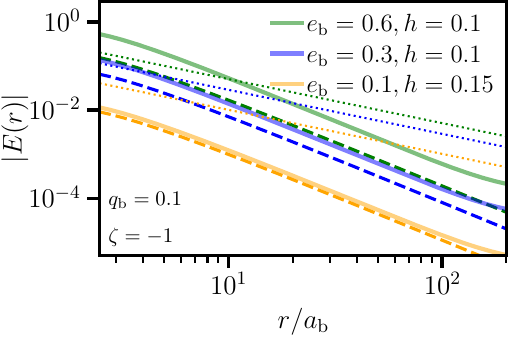}
    \caption{Disk eccentricity profiles $E(r)$ for a disk around an eccentric binary for three pairs of values of the binary eccentricity and the disk aspect ratio: $(\eb,h)=(0.6,0.1)$ (green lines), $(\eb,h)=(0.3,0.1)$ (blue lines), $(\eb,h)=(0.1,0.15)$ (orange lines). We plot numerical solutions (solid lines) and approximations given by Eq. (\ref{eq: forced approx example}) (dashed lines).
      The disk is a 2D locally isothermal disk with a density profile $\Sigma\propto r^{-1}$ and inner and outer radii $\rin=2.5\ab$ and $\rout=200\ab$, and the value of the binary mass ratio is $\qb=0.1$. For these, pressure-dominated systems, the forced disk eccentricity is approximately proportional to $r^{-2}$. 
      The eccentricity profiles for test particles $e_\mathrm{f}(r)$ are shown with dotted lines.
      }
    \label{fig: E(r) forced omega=0 eb variation}
\end{figure}

In Figure \ref{fig: E(r) forced omega=0 eb variation}, we plot numerical solutions for the eccentricity profile and the solutions given by Eq. (\ref{eq: forced approx example}) for three different values of $(\eb,h)$.
For all three values of the binary eccentricity, the disk eccentricity profile differs from the forced eccentricity of a test particle $e_\mathrm{f}\propto r^{-1}$ (Eq. (\ref{eq: forced ecc})). Instead, all three eccentricity profiles fall off 
with increasing radius as $r^{-2}$, as predicted. The solution given by Eq. (\ref{eq: forced approx example}) is closest to the numerical solutions for $\eb=0.1,h=0.15$, for which the value of $\rs/\rin$ is largest, rendering the pressure-dominated approximation more accurate.

Recent numerical simulations by \cite{Siwek+_2023} found that there's a range of circumbinary systems for which the CBD experiences forced eccentricity with an eccentricity profile significantly different from the one predicted by (Eq. (\ref{eq: forced ecc})). Instead, they find that $E\propto r^{-1.9}$ is a good fit for their results. This is in a good agreement with the analytic approximation for the eccentricity profile given by Eq. (\ref{eq: forced approx example}).

\section{Precessing Binary}\label{sec: 5}

Generally, eccentric binaries will undergo apsidal precession due to General Relativity or the influence of the disk itself. Here we consider the latter. Two-dimensional numerical calculations of CBDs around eccentric binaries show that the disk can induce apsidal precession of the binary occurring at a much faster rate than evolution of the other elements, e.g., semi-major axis and eccentricity \cite{Tiede_DOrazio_2024}. Hence, we use these results for the binary precession rate in terms of disk mass and binary eccentricity to study the response of the disk eccentricity and precession rate, including a generalization of resonant CBD eccentricity, in the presence of a precessing, eccentric binary.

\subsection{Binary Precession Due to a Massive Disk }\label{sec: 5a}

If the disk mass is sufficiently large, it will influence the binary’s orbital evolution. In general, the gravitational force exerted by the disk induces apsidal precession of the binary and drives changes in its separation and eccentricity, $\dot{a}_\mathrm{b}, \dot{e}_\mathrm{b}$.
We neglect the changes in $\ab$ and $\eb$ for two reasons. First, the secular changes (the binary precession rate) happen on faster timescales \citep{Tiede_DOrazio_2024} so we assume that $\ab$ and $\eb$ are constants on the disk precession timescales. Second, we assume that the gravitational potential caused by the disk mass is axisymmetric. An axisymmetric potential causes a change in the binary precession rate only.

We assume a finite disk with a central cavity of size $\rin$ and an outer radius $\rout$, and a power-law-density profile (Eq. (\ref{eq: sigma_zeta})). In Appendix \ref{appendix a}, we present the approximate analytical expressions for the gravitational potential inside such a disk ($\rin \leq r\leq \rout$) and within its central cavity ($r\leq \rin$).

Including the axisymmetric component of the disk's gravitational potential ($m=0$) results in two changes to the eccentricity equation compared to the one solved in Section \ref{sec: 4}. 
The first is a change of the binary precession rate in the $m=1$ contributions to the eccentricity equation (Eqs. \ref{eq: m=1 pot}, \ref{eq: gravitational forcing term w_bin}), which we discuss in \S \ref{subsection: Forced precession rate}. The second is an addition of the $m=0$ gravitational contribution of the disk potential in the eccentricity equation, which we discuss in \S \ref{subsection: Forced eccentricity profile} and Appendix \ref{appendix a}.

\subsection{Forced Precession Rate} \label{subsection: Forced precession rate}

The gravitational potential from the disk exerts a radial force on the binary causing its orbit to precess. In Appendix \ref{appendix a}, we calculate the precession rate by using an approximate expression for the gravitational potential inside the disk cavity to calculate a radial force on the orbit and the corresponding averaged binary precession rate. The result reads:
\begin{equation}\label{eq: average binary precession }
\begin{split}
\frac{\bomb}{\ob}=&-2\pi \qd \sqrt{1-\eb^2}\Bigg[3 a_1 \left(\frac{\rout}{\ab} \right)^{\zeta-1}\\
    & + a_2\left( \frac{\rout}{\ab}\right)^{\zeta-3}\left(10+\frac{15}{2}\eb^2\right) \Bigg],
    \end{split}
\end{equation}
where $a_1$ and $a_2$ are constants given by
\begin{subequations}
    \begin{equation}
        a_1= \frac{1}{4(1-\zeta)}\left[ 1-\left(\frac{\rin}{\rout}\right)^{\zeta-1}\right],
    \end{equation}
    \begin{equation}
        a_2= \frac{9}{64(3-\zeta)}\left[ 1-\left(\frac{\rin}{\rout}\right)^{\zeta-3}\right],
    \end{equation}
\end{subequations}
and where we have defined $\qd$ as a local disk-to-binary mass ratio \citep{Tiede_DOrazio_2024}:
\begin{equation}\label{eq: q_d}
    \qd=\frac{\Sigma_0 \ab^2}{\Mb}.
\end{equation}
As expected, the prefactor of Eq. (\ref{eq: average binary precession }), shows that increasing the value of the disk-to-binary mass ratio increases the value of the binary precession frequency. 

The exact mass of the disk depends on its size and the chosen density profile. For the sake of simplicity, we use $\qd$
as a measure of the influence of the disk mass on the CBD-binary system evolution. We note that the ratio of the total disk to binary mass can be several orders of magnitude larger than the parameters $\qd$. For example, for a $\Sigma_0=10^{-3}$, and $\rin=2\ab,\rout=100\ab$, $\qd=10^{-3}, \zeta=-1/2$, but $M_\mathrm{disk}/\Mb=4$. We use $\qd \approx 10^{-3}$ as the maximum value of the disk-to-binary mass ratio, as we expect the disk to be gravitationally unstable for $\qd \gtrsim 10^{-3}$ for standard $\alpha$-disks \citep{Tiede_DOrazio_2024}.

\begin{figure}[h!]
    \centering
    \includegraphics[scale=1]{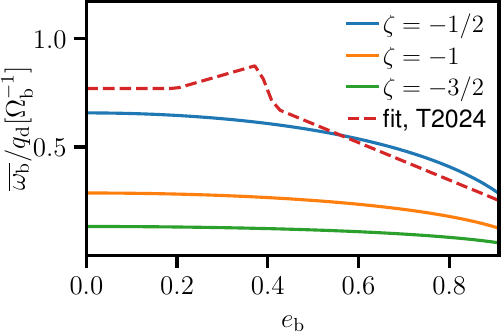}
    \caption{The binary precession rate, given by Eq. (\ref{eq: average binary precession }) (solid blue line). The fit for the binary precession rate obtained in hydrodynamical simulations by \cite{Tiede_DOrazio_2024} (dotted red line), which uses $\zeta (r\gg \ab) =0$. Disk inner and outer radii are $\rin=3\ab$ and $\rout=200\ab$. The value of the binary precession rate decreases with increasing value of the binary eccentricity, and is highest for $\zeta=-1/2$.}
    \label{fig: w0-qd forced precessing}
\end{figure}

In Figure \ref{fig: w0-qd forced precessing}, we plot the precession rate of the binary given by Eq. (\ref{eq: average binary precession }). The binary precession rate decreases with increasing binary eccentricity. In the same figure, we plot a fit for the binary apsidal precession frequency
obtained by \cite{Tiede_DOrazio_2024}.
Both our Eq. (\ref{eq: average binary precession }) and the T2024 fit show similar trends, except for a sharp change at $\eb\approx0.4$ in the T2024 fit, which is likely related to the change in cavity shape due to non-linear effects at this eccentricity.
The value of the binary precession frequency increases with increasing value of the power-law-density exponent $\zeta$, due to the mass of the disk being more concentrated close to the binary, thus exerting a greater force. Similarly, a decreasing value of the disk's inner radius increases the binary precession rate.

For the radial eccentricity distribution solution $E(r)$ to be time-independent, the disk precession frequency $\omega$ needs to be equal to the binary precession rate $\bomb$. 
Therefore, we set the disk precession frequency to be equal to that of the binary $\omega=\bomb$, and the disk precession frequency is identical to what is plotted in Figure \ref{fig: w0-qd forced precessing} for what follows.

\subsection{Forced Eccentricity Profile}\label{subsection: Forced eccentricity profile}

Next, we determine the eccentricity profile of the forced, precessing CBD mode, and show how the disk mass modifies the solution presented in Section \ref{sec: 4}. 

In addition to introducing the non-zero binary precession rate, computed from inserting Eq. (\ref{eq: average binary precession }) into Eq. (\ref{eq: m=1 pot}), we add to the quadrupole potential the (axisymmetric) contribution from the disk mass, so that $\Phi_{m=0} = \Phi_Q + \Phi_\mathrm{d}$ (Eq. (\ref{eq: potential binary m=0})). We use the approximate expression for the gravitational potential within the disk (Appendix \ref{appendix a}) to find the $m=0$ gravitational contribution to the eccentricity equation (Eq. (\ref{eq: gravitational forcing term for the eccentricty equation})).

\begin{figure}
\vspace{0pt}
$
\begin{array}{r}
\includegraphics[scale=0.91]{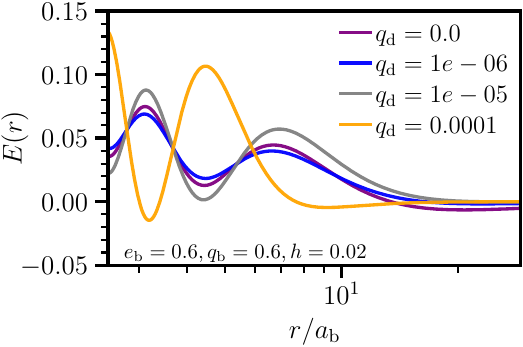} \\
\includegraphics[scale=0.91]{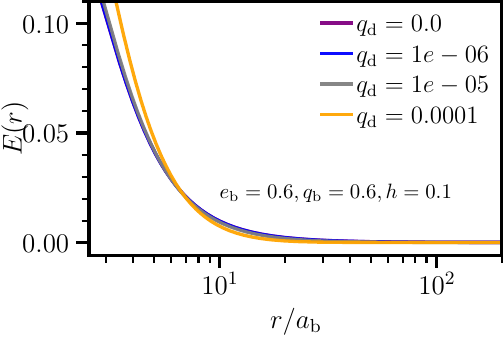}  \\
\end{array}$
\caption{Forced CBD mode eccentricity profiles $E(r)$ for three values of the disk-to-binary mass ratio; $\qd=0$ (solid purple line), $\qd= 10^{-6}$ (solid blue line), $\qd= 10^{-5}$ (solid grey line), and $\qd=10^{-4}$ (solid orange line). The values of the binary mass ratio and eccentricity are $\qb=\eb=0.6$, and the values of the disk aspect ratio are $h=0.02$ (top panel) and $h=0.1$ (bottom panel). The disk inner and outer radii are $\rin=2.5\ab$ and $\rout=200\ab$. The value of the density-power-law exponent is $\zeta=-1/2$. Increasing value of the disk-to-binary mass ratio changes the eccentricity profile significantly only for $\qd\gtrsim 10^{-4}$. }
\label{fig: E(r) two panel}
\end{figure}

We next describe solutions to our new eccentricity equation for a massive disk a round a precessing binary.
Figure \ref{fig: E(r) two panel} plots the forced disk eccentricity profile $E(r)$ for a 2D locally isothermal disk for four values of the disk-to-binary mass ratio and two values of the disk aspect ratio. 

In both panels it is clear that the disk mass has a minimal effect on the disk eccentricity profile, except for one case in the top panel where $q_d=10^{-4}$. The reason for this significant change in this one case, is a change in mode order above the ground mode, caused by the increased disk mass potential. Hence, we conclude that the effect of disk mass on eccentricity profile will be small unless $q_d$ is large enough to allow higher mode orders, which are more closely spaced. For chosen parameters $(\zeta=-1/2,\rin=2.5\ab,\rout=200\ab)$, the exact ratios of the disk to binary mass are $M_{\rm disk}/M_{\rm b}\approx\{1,0.1,0.01\}$ for $\qd=\{10^{-4},10^{-5},10^{-6}\}$.

\subsubsection{Resonant Disk Eccentricity}\label{subsection: Resonant eccentricity - massive disk}

Similarly to the non-precessing forced modes that we discussed in Section \ref{sec: 4}, we expect resonant disk eccentricity for systems for which the binary precession rate is equal to the precession frequency of the free CBD mode, i.e., $\omega=\bomb$. The values of various physical parameters also determine the number of radial nodes, and thus the order $n$, associated with such modes.

\begin{figure}
\vspace{0pt}
$
\begin{array}{r}
\includegraphics[scale=0.99]{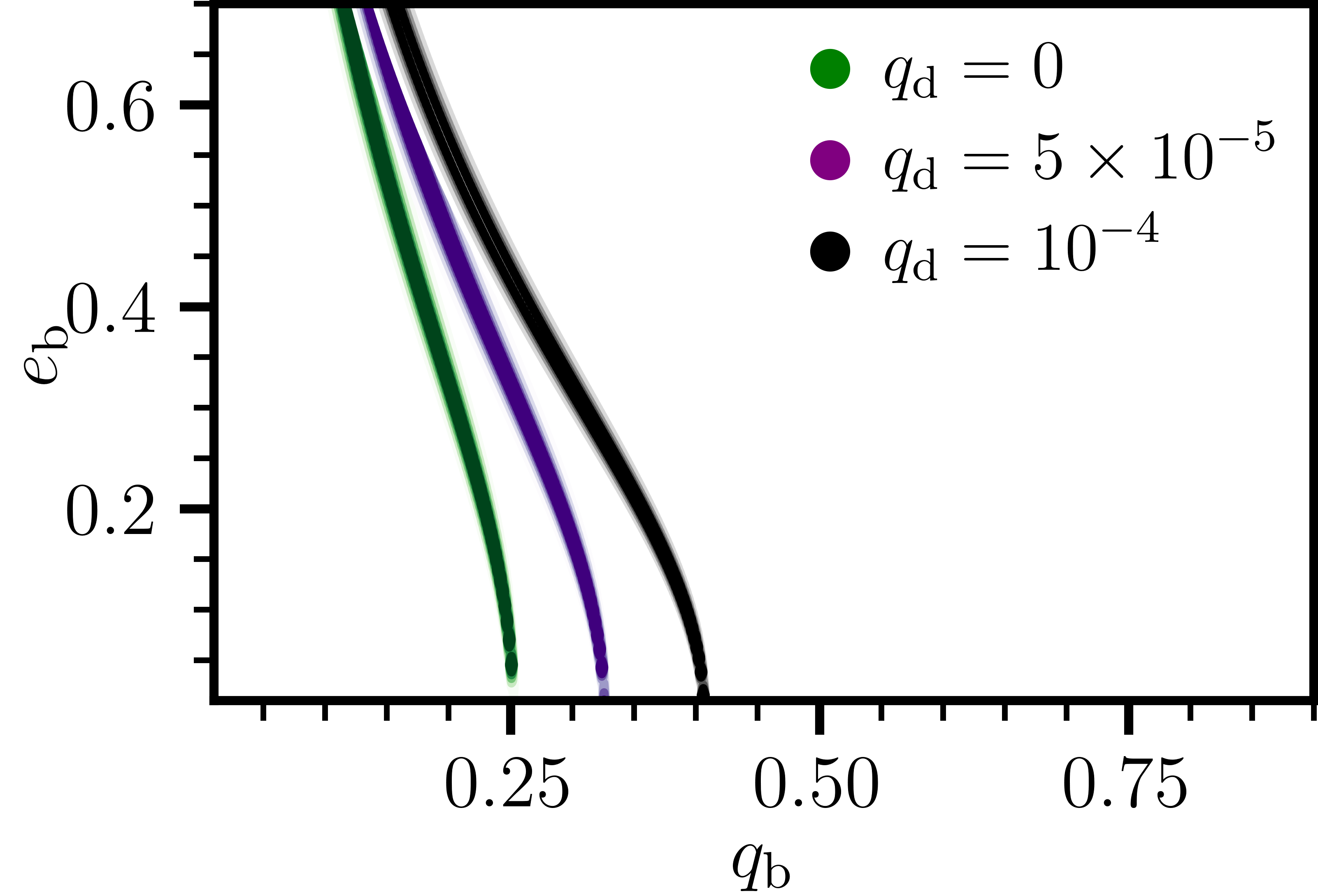} \\
\includegraphics[scale=0.99]{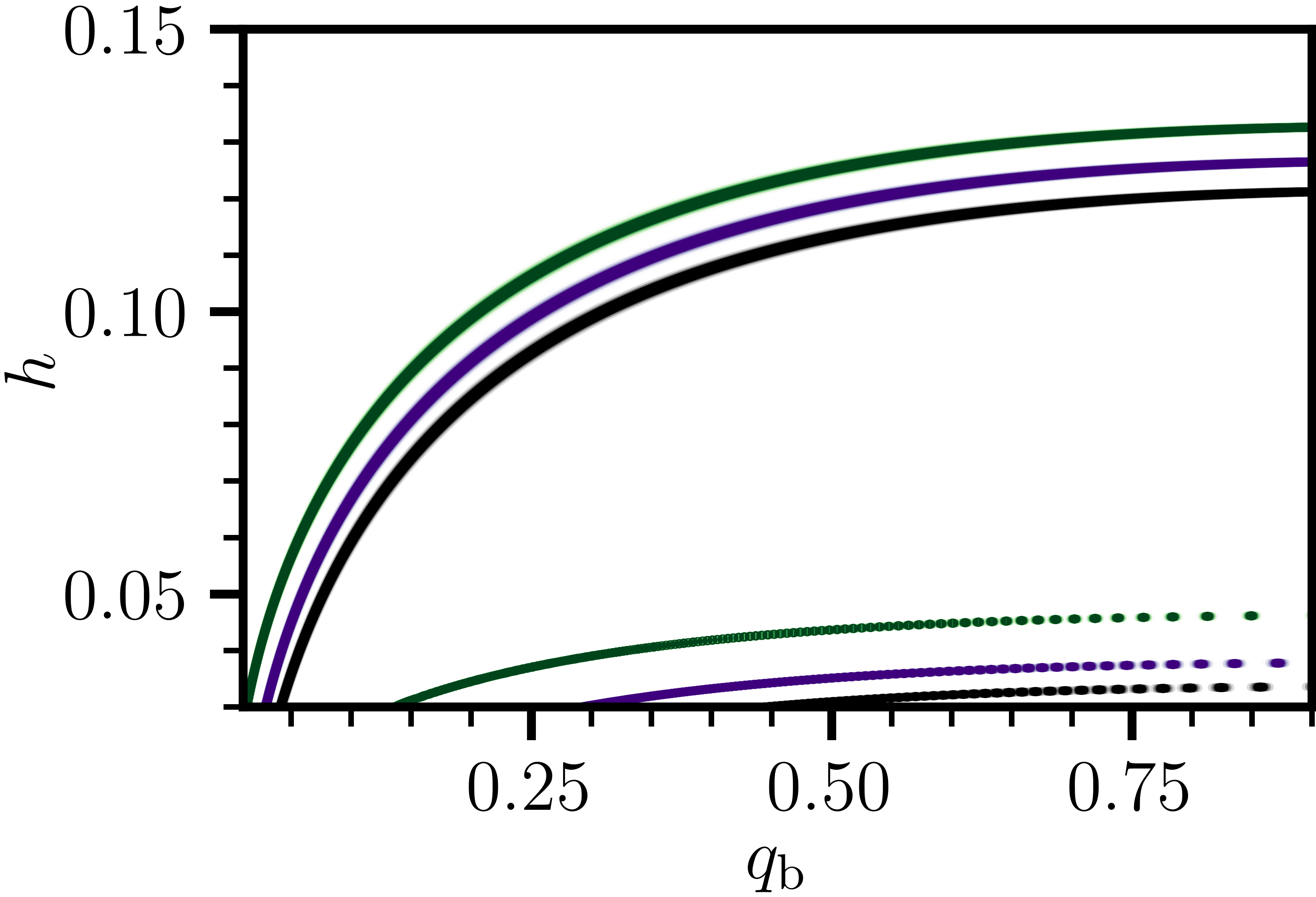}  \\
\includegraphics[scale=0.99]{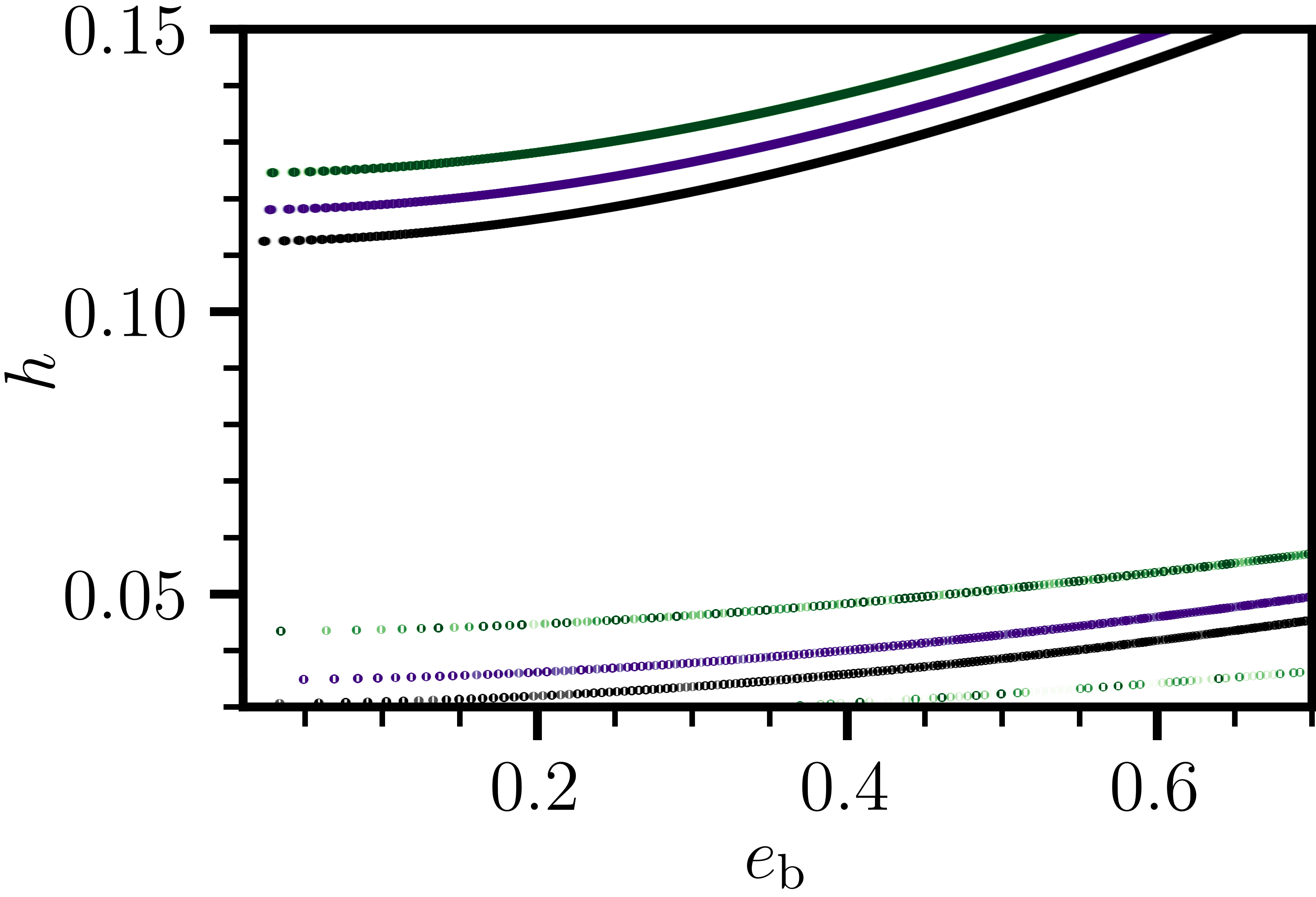}\\ 
\end{array}$
\caption{Contour plots of the forced disk eccentricity value at the disk inner edge $E(\rin)$ for varying values of the binary mass ratio and eccentricity with $h=0.1$ (panel 1); the binary mass ratio and the disk aspect ratio with $\eb=0.3$ (panel 2); the binary eccentricity and the disk aspect ratio with $\qb=0.9$ (panel 3). The disk is a 2D locally isothermal disk with a density profile $\Sigma=\Sigma_0 (r/\ab)^{-1/2}$, and inner and outer radii $\rin=2.5\ab$ and $\rout=200\ab$. Colored areas denote parameters for which the disk eccentricity at the inner radius is $E(\rin)\gtrsim 3$ for $\qd=0$ (green), $\qd=5 \times 10^{-5}$ (purple), and $\qd=10^{-4}$ (black).}
\label{fig: contours 2}
\end{figure}

In Figure \ref{fig: contours 2}, we show how the values of resonant parameters from Figure \ref{fig:contour plot 1} change for varying values of the disk-to-binary mass ratio, $q_d$. We show only the contours for which the value of the disk eccentricity at the disk inner radius is above $E(\rin)\gtrsim 3$.
Increasing the value of the disk-to-binary mass ratio increases the values of the binary mass ratio and the binary eccentricity at which resonant disk eccentricity occurs, while the opposite trend occurs with disk aspect ratio $h$. This trend occurs because the precession frequency of the forced mode increases with increasing $\qd$, and the precession frequency of the free mode needs to match that growth in order for a resonance to occur. Since the free mode precession frequency decreases with increasing $\qd$ (\S \ref{subsec: free modes}), the precession frequency growth necessary for a resonance needs to be achieved by either increasing positive quadrupole precession contribution or decreasing the negative pressure frequency contribution.

\begin{figure}[h!]
    \centering
    \includegraphics[scale=0.95]{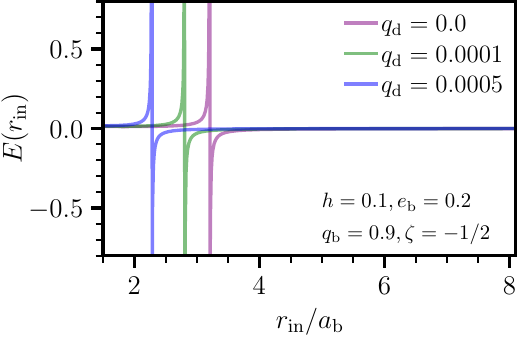}
    \caption{The value of the forced disk eccentricity at the disk inner radius for three values of the disk-to-binary mass ratio; $\qd=0$ (solid purple line), $\qd=10^{-4}$ (solid green line), $\qd=0.0005$ (solid blue line). The inner disk radius for which the resonant eccentricity values occur decreases with increasing value of the disk-to-binary mass ratio. The outer disk radius is $\rout=200\ab$.}
    \label{fig: E(in)-eb forced precessing}
\end{figure}

In addition to the three parameters ($\eb$, $\qb$, and $h$), the resonant disk eccentricity is also determined by the value of the disk inner radius $\rin$. 
In Figure \ref{fig: E(in)-eb forced precessing} we plot the forced eccentricity at the inner edge of the disk against inner disk radii values $1.5\ab\leq \rin \leq 8\ab$. Increasing the disk-to-binary mass ratio leads to a smaller value of the inner radius for which the value of the resonant disk eccentricity occurs. 

\section{Discussion and Conclusion}\label{sec: Discussion and Conclusion}

The gravitational potential of an eccentric, unequal-mass binary contains non-axisymmetric terms that cause a forced eccentricity in the circumbinary disk (CBD). In steady state, the two quantities that characterize the forced eccentric evolution of the CBD are its precession rate and the radial eccentricity profile. The precession rate of the forced disk eccentricity is equal to the precession rate of the binary. Since the results for the disk eccentricity profile depend on the binary precession rate, we consider the non-precessing and precessing binary cases separately. The precession rate of the binary, in turn, is determined by the gravitational potential of the disk. 

\paragraph{Non-precessing binary}  Often, the binary precession is set to zero, neglecting the disk mass and general relativity effects that can cause binary apsidal precession. For such systems, the precession rate of the forced CBD mode is zero (Section \ref{sec: 4}). The analytic results of \citet{p1} for non-precessing mode solutions allow us to preform an analytical study of forced non-precessing modes, too. We distinguish between the forced disk eccentricity distribution in pressure dominated and quadrupole dominated systems. For quadrupole dominated systems, such as thin disks around a near-equal mass binary, the forced disk eccentricity oscillates around the forced eccentricity of a test particle (\S \ref{subsec: quad dominated}, Fig. \ref{fig: E(r) forced omega=0 h variation}). The amplitude and wavelength of those oscillations is determined by the combination of physical parameters contained in the ratio $\rs/\rin = \sqrt{\wq/\wp}$ (Eq. \ref{eq: r_s definition}). For most combinations of values defining the system parameters, that amplitude is small compared to the value of the test particle forced eccentricity, and decreases with increasing $\rs/\rin$, i.e., $\wq \gg \wp$. In other words, decreasing pressure tends to damp the amplitude of the modes with respect to the test particle forced eccentricity.
However, in some quadrupole dominated systems, we can expect a resonant disk eccentricity value, as described in \cite{Lubow2022} and \S~\ref{subsection: Resonant Disk Eccentricity}. 
Using the solutions of \citet{p1} for power-law-density disks, Eq. (\ref{eq: zeta approximation non-precessing power law disks}) provides a simple predictive analytical formula for resonant behavior based on the necessary condition for the existence of non-trivial, non-precessing $(\omega=0)$ free modes.

In pressure-dominated disks, the forced disk eccentricity distribution deviates strongly from the forced eccentricity of a test particle (\S \ref{subsection: Pressure-dominated systems}). Specifically, it follows a power-law distribution in radius determined by the power-law-density exponent $\zeta$, the disk dimension, and the nature of the perturbation. 
For 2D locally isothermal disks, our prediction for the eccentricity profile of a forced CBD does not depend on $\zeta$, and is given as $E\propto r^{-2}$. This is in good agreement with the forced eccentricity profile $E\propto r^{-1.9}$ found  in the numerical simulations by \cite{Siwek+_2023}.
We also showed that disks with significant pressure support cannot satisfy the conditions for resonant  eccentricity excitation.

\paragraph{Precessing binary}  In Section \ref{sec: 5}, we consider disk massive enough to influence the CBD-binary system evolution.
We assume power-law-density disks, and find that the precession rate of the binary is proportional to the disk-to-binary mass ratio, and decreases with increasing value of the binary eccentricity (see Eq. \ref{eq: average binary precession rate}). This behavior is in agreement with the analysis and numerical simulations performed by \cite{Tiede_DOrazio_2024}, as shown in Figure \ref{fig: w0-qd forced precessing}. We find that the precession rate increases with increasing values of the power-law-density exponent $\zeta$, which we attribute to an increased total enclosed mass of the disk near the binary. We showed that the disk mass has a small effect on the disk eccentricity for disk-to-binary mass ratio $\qd\lesssim10^{-5}$, or roughly $M_{\rm disk}/M_{\rm b}=0.04$, and that the effect is stronger for higher-order modes (\S \ref{subsection: Forced eccentricity profile}).  

\paragraph{Precessing resonant eccentricity} Since the disk and binary precession rates are equal for a system in steady-state, a precessing binary changes the parameters that lead to resonant disk eccentricity growth, compared to the non-precessing case. We assumed that the binary precession is caused by the mass of the disk, and found that increasing the disk-to-binary mass ratio increases the resonance-supporting values of the binary mass ratio and eccentricity, and decreases resonance-supporting values of the disk aspect ratio (\S \ref{subsection: Resonant eccentricity - massive disk}). This is because increasing disk-to-binary mass ratio $\qd$ increases the disk precession frequency, and to make that higher frequency still be the free CBD-mode precession frequency, the pressure in the disk must decrease, or the quadrupole potential from the binary must increase.

We restricted the analysis to disks with a power-law-density profile and disk-mass induced binary precession, but a similar resonant behavior is expected for other density profiles and binary precession causes. For example, general relativity effects could cause apsidal binary precession, which would excite different free modes than considered here.  In the case of non-power-law disks, resonance could occur for thicker disks than were found not to support resonance here. This is because disks with a smoothly truncated cavity can support free modes even in the pressure-dominated, thicker disk regime (see, e.g., Figs. 7 and 13 in \cite{p1}).

\paragraph{Potential Implications for Disk Cavity Size} 
The implications of extreme disk eccentricity on the disk-binary system are unclear. Such large eccentricities have not been documented in numerical simulations and it is not immediately apparent how such resonances manifest in the disk structure, or how they depend on the details of the problem setup. Here, we offer one interpretation of resonant results in terms of the disk cavity size. Specifically, we posit that there is a maximum eccentricity for which the disk is steady and well defined.
To keep particles on bound orbits, the straightforward limit is $E(r)<1$. A more stringent requirement on the maximum disk eccentricity can be found by demanding that there is no orbit crossing of disk fluid particles \citep{Ogilvie_2001,Miranda_Rafikov_2018}. 
Where the disk eccentricity exceeds this limit (below a certain radius), the intersecting orbits could truncate or disrupt the disk. Figure \ref{fig: E(in)-eb forced precessing} demonstrates that 1) for a set of binary-and-disk parameters and binary precession frequency, the inner radius can be chosen such that a resonance occurs, and 2) that the eccentricity is larger than $E(r)>0.1$ near resonances. Our interpretation then is that as a disk diffuses inwards towards the binary, it will sweep through different inner disk radii until it reaches a radius where the disk eccentricity is resonantly excited, making possible the truncation of the disk at that radius. 

\begin{figure}[h!]\centering
    \includegraphics[scale=0.95]{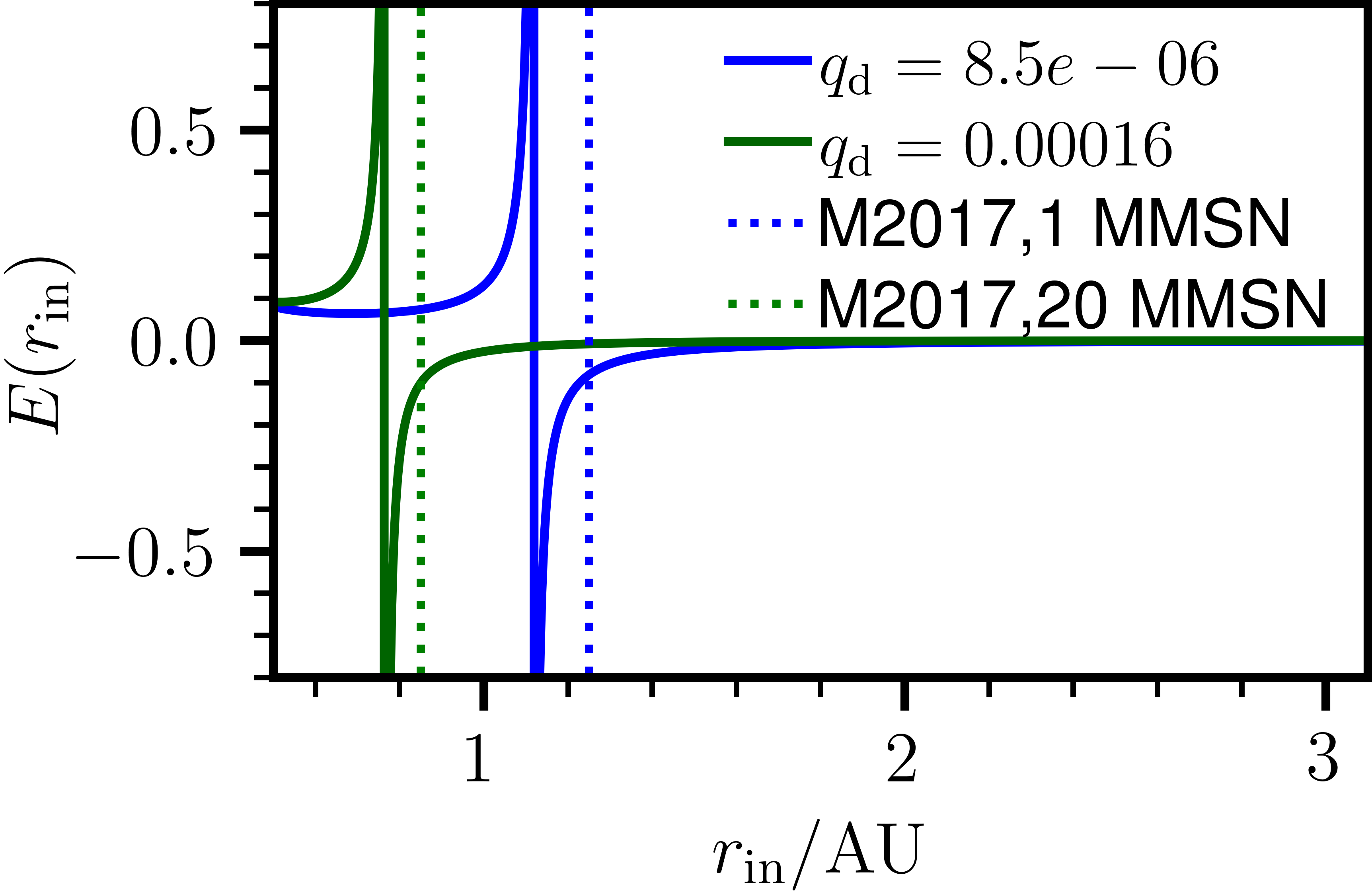}
    \caption{The value of the forced disk eccentricity at the disk inner radii for disk inner radius values in range $1.5\ab\leq \rin \leq 8.1\ab$, with $\ab=0.224 \mathrm{AU}$.
    for two values of the disk-to-binary mass ratio (solid lines). Estimates for the size of the disk central cavity, obtained from Figure 7a from \cite{Mutter_2017_a} are plotted in dotted lines. The binary precession rate is fixed at zero to match the setup of \cite{Mutter_2017_a}.
    }\label{fig:placeholder - e(in)-rin}
\end{figure}

As a case study, we consider how such a resonantly truncated inner radius changes with changing disk mass and compare to the results of numerical hydrodynamical simulations by \cite{Mutter_2017_a}. The inner radius needed for resonance decreases as the value of the disk-to-binary mass ratio increases. \cite{Mutter_2017_a} performed hydrodynamics simulations of disks with self-gravity around binaries (cutting out the binary from the simulation domain) and found that the cavity size decreases with increasing disk mass. We provide a rough estimate to enable a comparison between their results and ours. We use system parameters from \cite{Mutter_2017_a} for Kepler-16: $\qb=0.294,\eb=0.159, h=0.05$. In Figure \ref{fig:placeholder - e(in)-rin}, we plot as vertical dotted lines our estimate of the cavity size \citep[Figure 7a, ][]{Mutter_2017_a} for two values of the disk mass: a 1$\times$ and 20$\times$ minimum mass solar nebula (MMSN). In Figure \ref{fig:placeholder - e(in)-rin}, we plot the disk eccentricity at the disk inner radius as a function of the value of the inner disk radius (See Section \ref{sec: 5} and Figure \ref{fig: E(in)-eb forced precessing} for details on how we obtain it). We plot results for two values of the disk-to-binary mass ratios, chosen so that the disk mass is close to 1 and 20 MMSN, which we take to be $\textit{MMSN}=0.01 M_\odot$, and we assume $\rout=10\ab$ and $\zeta=-1/2$. In Figure \ref{fig:placeholder - e(in)-rin}, unlike in Section \ref{sec: 5}, we assume that the disk precession rate is zero to better present the binary system in \cite{Mutter_2017_a}, which is on a fixed non-precessing orbit. With these simplifications, the radii where our computed disk eccentricity exceeds $E\sim0.1$ match quite well with values of inner disk radii obtained by \cite{Mutter_2017_a}.
The one-to-one comparison between the results is not possible here, as our disk density model is more simplistic than what \cite{Mutter_2017_a} obtained in simulations. Moreover, the disk in \cite{Mutter_2017_a} exhibits complicated evolution, such as evidence for higher order and multiple mode structure, with eccentricity profiles exhibiting multiple nodes and two distinct precession frequencies identified in different regions of the disk. While intriguing, the details of the disk evolution found in \cite{Mutter_2017_a} are beyond the scope of this work. A detailed investigation of these features would warrant further investigation and be better developed with a suite of tailored simulations.

Importantly, the values of inner disk radii obtained by \cite{Mutter_2017_a} for all three systems they studied (Kepler-16, Kepler-35, and Kepler-36) follow the trend we predicted of the inner disk radius decreasing with decreasing disk mass.

\paragraph{A Resonant Origin for the Cavity Radius} 
The standard picture for what sets the cavity size in a CBD is not entirely agreed upon, but has been considered in terms of ($m>1$) Linblad torques as well as the loss of stable particle orbits in regions too near the binary \citep{RudakPac1981, Artymowicz_Lubow_1994, Dan-gaps, Mahesh_plus_2024}. Our above picture suggests another possible dynamical origin for the cavity size for eccentric, non-equal-mass binaries.
If the gas near the inner edge of a CBD diffuses inward slowly compared with the time required for the eccentric response to adjust, the disk will sample a sequence of inner radii, each associated with a different spectrum of free eccentric modes. In this situation, inward motion of the cavity edge may continue until one of these modes approaches resonance with the binary forcing frequency, $\omega_{n}(\rin)\simeq \bar{\omega}_{\rm b}$. 
Among the possible resonances, the fundamental mode is the natural first candidate to regulate the cavity edge. It has the lowest free precession frequency and a node-free, global structure, making it the mode most naturally matched to the slow, large-scale forcing from a precessing binary. Its smooth radial profile should also maximize the forcing overlap and reduce the cancellations expected for higher-order modes, while making it less susceptible to damping by viscosity, shocks, or turbulent diffusion. Near the condition $\omega_{0}(\rin)\simeq\bar{\omega}_{\rm b}$, the forced eccentric response could therefore be amplified strongly enough to enhance the non-axisymmetric disk--binary interaction and increase the angular-momentum flux deposited near the cavity edge. This feedback may balance, or arrest, the inward diffusive spreading of the disk, selecting an inner radius close to the resonant value, provided that this resonant inner radius is greater than a cavity set by Linblad torques or particle orbit instabilities. This interpretation goes beyond the linear boundary-value problem, which identifies the resonant radii but does not by itself determine whether the disk edge evolves toward them. It is nevertheless a sensible conjecture
that could be tested by determining whether disks stall near radii satisfying $\omega_{0}(\rin)\simeq\bar{\omega}_{\rm b}$, or more generally $\omega_{n}(\rin)\simeq\bar{\omega}_{\rm b}$, 
and whether the associated resonant response and orbit crossing indeed truncates the disk.

\paragraph{Comparison to other works } Previous studies have shown that disks around eccentric binaries exhibit a precessing eccentricity distribution, with the precession rate being different from that of the binary \citep{Siwek+_2023,Tiede_DOrazio_2024}. This contrasts the simple picture of the forced disk eccentricity described in Section \ref{sec: 4} and Section \ref{sec: 5}. One possible explanation for the discrepancy might be the presence of multiple eccentric modes in the disk. \cite{Siwek+_2023} has found that the angle of the forced eccentric modes with respect to the binary apsidal orientation differs from zero, and depends on the value of the binary mass ratio and eccentricity. We have found no such locking angle dependency on system parameters because we have neglected bulk viscosity. The bulk viscosity can be accounted for as an imaginary component of the adiabatic constant $\gamma$, leading to the complex eccentricity $E(r)$ \citep{GO2006,Lubow2022}. The angle of the complex forced eccentricity is then given by $\arctan(\Im{E}/\Re{E})$, which is explored in \citet{Lubow2022}. 
Another consequence of neglecting the bulk disk viscosity is the lack of eccentricity damping. When bulk viscosity is accounted for, it damps the magnitude of the disk eccentricity, most notably for resonant systems, resulting in a finite eccentricity even at resonance-inducing parameters \citep{Lubow2022}.

\paragraph{Additional non-axisymmetric gravitational potential components}
The present analysis focuses on the secular $m=1$ component of the binary potential because it couples directly to the eccentric, $m=1$, disk response described by $E(r)$. Higher-order non-axisymmetric components, such as $m=2$, are dynamically important for tidal torques, spiral-wave excitation, and cavity formation, but their coupling to disk eccentricity is indirect or nonlinear in the present framework. Extending the analysis to include these components would require a treatment of additional azimuthal perturbations and their coupling to the eccentric mode, and is therefore beyond the scope of this work.
We have assumed the axial symmetry and time-independence of the gravitational potential of the disk. Although this is true for the reference state of the disk, the eccentric perturbations disturb the symmetry of the density distribution and the gravitational potential from that density. We assume that the values of the disk mass and eccentricity are low enough to make the correction to the axisymmetric potential minimal. However, for more massive and eccentric disks, a more careful analysis of the perturbed gravitational potential might be needed. We have also neglected the impact of streams, whose time-dependence could be relevant for the disk-mass potential, and very likely important for eccentricity growth through angular momentum and energy deposition in the disk \citep[e.g., ][]{Shi+_2012}.

While we argue that the perturbation theory methods discussed here continue to be useful in gaining insight into binary-disk dynamics and in revealing possible new behaviors, such as eccentric disk resonances, their connection to nonlinear hydrodynamical simulations has still not been entirely elucidated. This is in part due to the complicated non-linear response of the disk near the binary, e.g., accretion streams and shocks.
Future work could aim to bridge this gap with pointed, numerical experiments that can reproduce and extend the perturbative approach, and so provide further insight towards the role of free, forced, and resonant eccentricity evolution in CBDs. Such a program stands to enrich our understanding of the dynamics underlying the orbital evolution and observational signatures of accreting binaries.

\acknowledgments
We thank Alexander Dittmann and Steve Lubow for useful discussions.
M.G. and D.J.D. acknowledge support from the Danish Independent Research Fund through the Sapere Aude Starting Grant No. 121587. The research leading to this work received funding from the Independent Research Fund Denmark via grant ID 10.46540/3103-00205B. M.E.P. gratefully acknowledges the hospitality of the Institute for Advanced Study, where part of this work was carried out. DJD acknowledges support from the STScI Director's Discretionary fund. The Tycho supercomputer hosted at the SCIENCE HPC center at the University of Copenhagen was used to support this work.

\bibliographystyle{apj} 
\bibliography{refs}

@ARTICLE{RudakPac1981,
       author = {{Rudak}, B. and {Paczynski}, B.},
        title = "{Outer excretion disk around a close binary}",
      journal = {\actaa},
         year = 1981,
        month = jan,
       volume = {31},
       number = {1},
        pages = {13-24},
       adsurl = {https://ui.adsabs.harvard.edu/abs/1981AcA....31...13R}
}

@ARTICLE{Franchini_eccSG_2024,
       author = {{Franchini}, Alessia and {Prato}, Alessandra and {Longarini}, Cristiano and {Sesana}, Alberto},
        title = "{The behaviour of eccentric sub-pc massive black hole binaries embedded in massive discs}",
      journal = {\aap},
         year = 2024,
        month = aug,
       volume = {688},
          eid = {A174},
        pages = {A174},
          doi = {10.1051/0004-6361/202449402},
archivePrefix = {arXiv},
       eprint = {2402.00938},
 primaryClass = {astro-ph.HE},
       adsurl = {https://ui.adsabs.harvard.edu/abs/2024A&A...688A.174F}
}

@ARTICLE{Tiwari_RMHD_I+2025,
       author = {{Tiwari}, Vishal and {Chan}, Chi-Ho and {Bogdanovi{\'c}}, Tamara and {Jiang}, Yan-Fei and {Davis}, Shane W. and {Ferrel}, Simon},
        title = "{Radiation Magnetohydrodynamic Simulation of Sub-Eddington Circumbinary Disk around an Equal-mass Massive Black Hole Binary}",
      journal = {\apj},
         year = 2025,
        month = jun,
       volume = {986},
       number = {2},
          eid = {158},
        pages = {158},
          doi = {10.3847/1538-4357/add408},
archivePrefix = {arXiv},
       eprint = {2502.18584},
 primaryClass = {astro-ph.HE},
       adsurl = {https://ui.adsabs.harvard.edu/abs/2025ApJ...986..158T}
}

@ARTICLE{ClyburnZrake:2026,
       author = {{Clyburn}, Madeline and {Zrake}, Jonathan},
        title = "{Unequal Mass Binary Evolution Driven by High Mach Circumbinary Disks}",
      journal = {\mnras},
         year = 2026,
        month = mar,
          doi = {10.1093/mnras/stag567},
       adsurl = {https://ui.adsabs.harvard.edu/abs/2026MNRAS.tmp..552C}
}

@ARTICLE{Siwek_orbevo+2023,
       author = {{Siwek}, Magdalena and {Weinberger}, Rainer and {Hernquist}, Lars},
        title = "{Orbital evolution of binaries in circumbinary discs}",
      journal = {\mnras},
         year = 2023,
        month = jun,
       volume = {522},
       number = {2},
        pages = {2707-2717},
          doi = {10.1093/mnras/stad1131},
archivePrefix = {arXiv},
       eprint = {2302.01785},
 primaryClass = {astro-ph.HE},
       adsurl = {https://ui.adsabs.harvard.edu/abs/2023MNRAS.522.2707S}
}

@ARTICLE{Siwek_PTAsorbevo+2024,
       author = {{Siwek}, Magdalena and {Kelley}, Luke Zoltan and {Hernquist}, Lars},
        title = "{Signatures of circumbinary disc dynamics in multimessenger population studies of massive black hole binaries}",
      journal = {\mnras},
         year = 2024,
        month = nov,
       volume = {534},
       number = {3},
        pages = {2609-2620},
          doi = {10.1093/mnras/stae2251},
archivePrefix = {arXiv},
       eprint = {2403.08871},
 primaryClass = {astro-ph.HE},
       adsurl = {https://ui.adsabs.harvard.edu/abs/2024MNRAS.534.2609S}
}

@ARTICLE{DOrazio_binlite+2024,
       author = {{D'Orazio}, Daniel J. and {Duffell}, Paul C. and {Tiede}, Christopher},
        title = "{Fast Methods for Computing Photometric Variability of Eccentric Binaries: Boosting, Lensing, and Variable Accretion}",
      journal = {\apj},
         year = 2024,
        month = dec,
       volume = {977},
       number = {2},
          eid = {244},
        pages = {244},
          doi = {10.3847/1538-4357/ad938b},
archivePrefix = {arXiv},
       eprint = {2403.05629},
 primaryClass = {astro-ph.HE},
       adsurl = {https://ui.adsabs.harvard.edu/abs/2024ApJ...977..244D}
}

@ARTICLE{Tofflemire_DQTau+2017,
       author = {{Tofflemire}, Benjamin M. and {Mathieu}, Robert D. and {Ardila}, David R. and {Akeson}, Rachel L. and {Ciardi}, David R. and {Johns-Krull}, Christopher and {Herczeg}, Gregory J. and {Quijano-Vodniza}, Alberto},
        title = "{Accretion and Magnetic Reconnection in the Classical T Tauri Binary DQ Tau}",
      journal = {\apj},
         year = 2017,
        month = jan,
       volume = {835},
       number = {1},
          eid = {8},
        pages = {8},
          doi = {10.3847/1538-4357/835/1/8},
archivePrefix = {arXiv},
       eprint = {1612.02431},
 primaryClass = {astro-ph.SR},
       adsurl = {https://ui.adsabs.harvard.edu/abs/2017ApJ...835....8T}
}

@ARTICLE{MurrayDuffell:2025,
       author = {{Murray}, Allen R. and {Duffell}, Paul C.},
        title = "{Accreting Binary Eccentricities Follow Predicted Equilibrium Values}",
      journal = {\apj},
         year = 2025,
        month = apr,
       volume = {982},
       number = {2},
          eid = {113},
        pages = {113},
          doi = {10.3847/1538-4357/adb96e},
archivePrefix = {arXiv},
       eprint = {2411.13489},
 primaryClass = {astro-ph.SR},
       adsurl = {https://ui.adsabs.harvard.edu/abs/2025ApJ...982..113M}
}

@ARTICLE{2023_Lai_Munoz_review,
       author = {{Lai}, Dong and {Mu{\~n}oz}, Diego J.},
        title = "{Circumbinary Accretion: From Binary Stars to Massive Binary Black Holes}",
      journal = {\araa},
         year = 2023,
        month = aug,
       volume = {61},
        pages = {517-560},
          doi = {10.1146/annurev-astro-052622-022933},
archivePrefix = {arXiv},
       eprint = {2211.00028},
 primaryClass = {astro-ph.HE},
       adsurl = {https://ui.adsabs.harvard.edu/abs/2023ARA&A..61..517L}
}

@ARTICLE{2024_Cocchiararo,
       author = {{Cocchiararo}, F. and {Franchini}, A. and {Lupi}, A. and {Sesana}, A.},
        title = "{Electromagnetic signatures from accreting massive black hole binaries in time domain photometric surveys}",
      journal = {\aap},
         year = 2024,
        month = nov,
       volume = {691},
          eid = {A250},
        pages = {A250},
          doi = {10.1051/0004-6361/202449598},
archivePrefix = {arXiv},
       eprint = {2402.05175},
 primaryClass = {astro-ph.HE},
       adsurl = {https://ui.adsabs.harvard.edu/abs/2024A&A...691A.250C}
}

@ARTICLE{2007_hayasaki,
       author = {{Hayasaki}, Kimitake and {Mineshige}, Shin and {Sudou}, Hiroshi},
        title = "{Binary Black Hole Accretion Flows in Merged Galactic Nuclei}",
      journal = {\pasj},
         year = 2007,
        month = apr,
       volume = {59},
        pages = {427-441},
          doi = {10.1093/pasj/59.2.427},
archivePrefix = {arXiv},
       eprint = {astro-ph/0609144},
 primaryClass = {astro-ph},
       adsurl = {https://ui.adsabs.harvard.edu/abs/2007PASJ...59..427H}
}

@ARTICLE{1986_Boss,
       author = {{Boss}, Alan P.},
        title = "{Protostellar Formation in Rotating Interstellar Clouds. V. Nonisothermal Collapse and Fragmentation}",
      journal = {\apjs},
         year = 1986,
        month = nov,
       volume = {62},
        pages = {519},
          doi = {10.1086/191150},
       adsurl = {https://ui.adsabs.harvard.edu/abs/1986ApJS...62..519B}
}

@BOOK{Murray_Dermott_1999,
       author = {{Murray}, Carl D. and {Dermott}, Stanley F.},
        title = "{Solar System Dynamics}",
         year = 1999,
          doi = {10.1017/CBO9781139174817},
       adsurl = {https://ui.adsabs.harvard.edu/abs/1999ssd..book.....M},
      publisher={Cambridge University Press}
}

@ARTICLE{Tiede_DOrazio_2024,
       author = {{Tiede}, Christopher and {D'Orazio}, Daniel J. and {Zwick}, Lorenz and {Duffell}, Paul C.},
        title = "{Disk-induced Binary Precession: Implications for Dynamics and Multimessenger Observations of Black Hole Binaries}",
      journal = {\apj},
         year = 2024,
        month = mar,
       volume = {964},
       number = {1},
          eid = {46},
        pages = {46},
          doi = {10.3847/1538-4357/ad2613},
archivePrefix = {arXiv},
       eprint = {2312.01805},
 primaryClass = {astro-ph.HE},
       adsurl = {https://ui.adsabs.harvard.edu/abs/2024ApJ...964...46T}
}

@ARTICLE{2023MNRAS.523.4353E,
       author = {{Elsender}, Daniel and {Bate}, Matthew R. and {Lakeland}, Ben S. and {Jensen}, Eric L.~N. and {Lubow}, Stephen H.},
        title = "{On the frequencies of circumbinary discs in protostellar systems}",
      journal = {\mnras},
         year = 2023,
        month = aug,
       volume = {523},
       number = {3},
        pages = {4353-4364},
          doi = {10.1093/mnras/stad1695},
archivePrefix = {arXiv},
       eprint = {2306.06035},
 primaryClass = {astro-ph.EP},
       adsurl = {https://ui.adsabs.harvard.edu/abs/2023MNRAS.523.4353E}
}

@ARTICLE{Moriwaki_Nakagawa_2004,
       author = {{Moriwaki}, Kazumasa and {Nakagawa}, Yoshitsugu},
        title = "{A Planetesimal Accretion Zone in a Circumbinary Disk}",
      journal = {\apj},
         year = 2004,
        month = jul,
       volume = {609},
       number = {2},
        pages = {1065-1070},
          doi = {10.1086/421342},
       adsurl = {https://ui.adsabs.harvard.edu/abs/2004ApJ...609.1065M}
}

@ARTICLE{Mahesh_plus_2024,
       author = {{Mahesh}, Siddharth and {McWilliams}, Sean T. and {Pirog}, Michal},
        title = "{Analytical and Numerical Analysis of Circumbinary Disk Dynamics. I. Coplanar Systems}",
      journal = {\apj},
         year = 2024,
        month = sep,
       volume = {973},
       number = {1},
          eid = {18},
        pages = {18},
          doi = {10.3847/1538-4357/ad6149},
archivePrefix = {arXiv},
       eprint = {2305.01533},
 primaryClass = {astro-ph.SR},
       adsurl = {https://ui.adsabs.harvard.edu/abs/2024ApJ...973...18M}
}

@ARTICLE{Hure_Pierens_2005,
       author = {{Hur{\'e}}, Jean-Marc and {Pierens}, Arnaud},
        title = "{Accurate Numerical Potential and Field in Razor-thin, Axisymmetric Disks}",
      journal = {\apj},
         year = 2005,
        month = may,
       volume = {624},
       number = {1},
        pages = {289-294},
          doi = {10.1086/428769},
archivePrefix = {arXiv},
       eprint = {astro-ph/0501185},
 primaryClass = {astro-ph},
       adsurl = {https://ui.adsabs.harvard.edu/abs/2005ApJ...624..289H}
}

@ARTICLE{Tremaine_2001,
       author = {{Tremaine}, Scott},
        title = "{Slow Modes in Keplerian Disks}",
      journal = {\aj},
         year = 2001,
        month = mar,
       volume = {121},
       number = {3},
        pages = {1776-1789},
          doi = {10.1086/319398},
archivePrefix = {arXiv},
       eprint = {astro-ph/0011571},
 primaryClass = {astro-ph},
       adsurl = {https://ui.adsabs.harvard.edu/abs/2001AJ....121.1776T}
}

@ARTICLE{2022_gutierrez,
       author = {{Guti{\'e}rrez}, Eduardo M. and {Combi}, Luciano and {Noble}, Scott C. and {Campanelli}, Manuela and {Krolik}, Julian H. and {L{\'o}pez Armengol}, Federico and {Garc{\'\i}a}, Federico},
        title = "{Electromagnetic Signatures from Supermassive Binary Black Holes Approaching Merger}",
      journal = {\apj},
         year = 2022,
        month = apr,
       volume = {928},
       number = {2},
          eid = {137},
        pages = {137},
          doi = {10.3847/1538-4357/ac56de},
archivePrefix = {arXiv},
       eprint = {2112.09773},
 primaryClass = {astro-ph.HE},
       adsurl = {https://ui.adsabs.harvard.edu/abs/2022ApJ...928..137G}
}

@article{Izzard,
	author = {Izzard, Robert G and Jermyn, Adam S},
	doi = {10.1093/mnras/stac2899},
	eprint = {https://academic.oup.com/mnras/article-pdf/521/1/35/49406979/stac2899.pdf},
	issn = {0035-8711},
	journal = {Monthly Notices of the Royal Astronomical Society},
	month = {05},
	number = {1},
	pages = {35-50},
	title = {Circumbinary discs for stellar population models},
	url = {https://doi.org/10.1093/mnras/stac2899},
	volume = {521},
	year = {2023}}

@ARTICLE{2026_Unger,
       author = {{Unger}, Lotem and {Grichener}, Aldana and {Soker}, Noam},
        title = "{Circumbinary Disks in Post Common Envelope Binary Systems with Compact Objects}",
      journal = {\apj},
         year = 2026,
        month = feb,
       volume = {998},
       number = {1},
          eid = {79},
        pages = {79},
          doi = {10.3847/1538-4357/ae3241},
archivePrefix = {arXiv},
       eprint = {2411.15652},
 primaryClass = {astro-ph.SR},
       adsurl = {https://ui.adsabs.harvard.edu/abs/2026ApJ...998...79U}
}

@ARTICLE{Hure_Hersant_2007,
       author = {{Hur{\'e}}, J. -M. and {Hersant}, F.},
        title = "{A new equation for the mid-plane potential of power law disks}",
      journal = {\aap},
         year = 2007,
        month = jun,
       volume = {467},
       number = {3},
        pages = {907-910},
          doi = {10.1051/0004-6361:20077132},
archivePrefix = {arXiv},
       eprint = {astro-ph/0703603},
 primaryClass = {astro-ph},
       adsurl = {https://ui.adsabs.harvard.edu/abs/2007A&A...467..907H}
}

@ARTICLE{Hure_Pelat_2007,
       author = {{Hur{\'e}}, J. -M. and {Pelat}, D. and {Pierens}, A.},
        title = "{Generation of potential/surface density pairs in flat disks. Power law distributions}",
      journal = {\aap},
         year = 2007,
        month = nov,
       volume = {475},
       number = {2},
        pages = {401-407},
          doi = {10.1051/0004-6361:20066808},
archivePrefix = {arXiv},
       eprint = {0706.3616},
 primaryClass = {astro-ph},
       adsurl = {https://ui.adsabs.harvard.edu/abs/2007A&A...475..401H}
}

@ARTICLE{Hure_Hersant_2008,
       author = {{Hur{\'e}}, J. -M. and {Hersant}, F. and {Carreau}, C. and {Busset}, J. -P.},
        title = "{A new equation for the mid-plane potential of power law discs. II. Exact solutions and approximate formulae}",
      journal = {\aap},
         year = 2008,
        month = nov,
       volume = {490},
       number = {2},
        pages = {477-486},
          doi = {10.1051/0004-6361:200809682},
archivePrefix = {arXiv},
       eprint = {0808.1634},
 primaryClass = {astro-ph},
       adsurl = {https://ui.adsabs.harvard.edu/abs/2008A&A...490..477H}
}

@ARTICLE{Mutter_2017_a,
       author = {{Mutter}, Matthew M. and {Pierens}, Arnaud and {Nelson}, Richard P.},
        title = "{The role of disc self-gravity in circumbinary planet systems - I. Disc structure and evolution}",
      journal = {\mnras},
         year = 2017,
        month = mar,
       volume = {465},
       number = {4},
        pages = {4735-4752},
          doi = {10.1093/mnras/stw2768},
archivePrefix = {arXiv},
       eprint = {1610.07811},
 primaryClass = {astro-ph.EP},
       adsurl = {https://ui.adsabs.harvard.edu/abs/2017MNRAS.465.4735M}
}

@ARTICLE{Ragusa+_2020,
       author = {{Ragusa}, Enrico and {Alexander}, Richard and {Calcino}, Josh and {Hirsh}, Kieran and {Price}, Daniel J.},
        title = "{The evolution of large cavities and disc eccentricity in circumbinary discs}",
      journal = {\mnras},
         year = 2020,
        month = dec,
       volume = {499},
       number = {3},
        pages = {3362-3380},
          doi = {10.1093/mnras/staa2954},
archivePrefix = {arXiv},
       eprint = {2009.10738},
 primaryClass = {astro-ph.EP},
       adsurl = {https://ui.adsabs.harvard.edu/abs/2020MNRAS.499.3362R}
}

@ARTICLE{Valli+2024,
       author = {{Valli}, Ruggero and {Tiede}, Christopher and {Vigna-G{\'o}mez}, Alejandro and {Cuadra}, Jorge and {Siwek}, Magdalena and {Ma}, Jing-Ze and {D'Orazio}, Daniel J. and {Zrake}, Jonathan and {de Mink}, Selma E.},
        title = "{Long-term evolution of binary orbits induced by circumbinary disks}",
      journal = {\aap},
         year = 2024,
        month = aug,
       volume = {688},
          eid = {A128},
        pages = {A128},
          doi = {10.1051/0004-6361/202449421},
archivePrefix = {arXiv},
       eprint = {2401.17355},
 primaryClass = {astro-ph.HE},
       adsurl = {https://ui.adsabs.harvard.edu/abs/2024A&A...688A.128V}
}

@ARTICLE{DOrazioCharisi:2023,
       author = {{D'Orazio}, Daniel J. and {Charisi}, Maria},
        title = "{Observational Signatures of Supermassive Black Hole Binaries}",
      journal = {arXiv e-prints},
         year = 2023,
        month = oct,
          eid = {arXiv:2310.16896},
        pages = {arXiv:2310.16896},
          doi = {10.48550/arXiv.2310.16896},
archivePrefix = {arXiv},
       eprint = {2310.16896},
 primaryClass = {astro-ph.HE},
       adsurl = {https://ui.adsabs.harvard.edu/abs/2023arXiv231016896D}
}

@ARTICLE{KITP_CC:2024,
       author = {{Duffell}, Paul C. and {Dittmann}, Alexander J. and {D'Orazio}, Daniel J. and {Franchini}, Alessia and {Kratter}, Kaitlin M. and {Penzlin}, Anna B.~T. and {Ragusa}, Enrico and {Siwek}, Magdalena and {Tiede}, Christopher and {Wang}, Haiyang and {Zrake}, Jonathan and {Dempsey}, Adam M. and {Haiman}, Zoltan and {Lupi}, Alessandro and {Pirog}, Michal and {Ryan}, Geoffrey},
        title = "{The Santa Barbara Binary‑disk Code Comparison}",
      journal = {\apj},
         year = 2024,
        month = aug,
       volume = {970},
       number = {2},
          eid = {156},
        pages = {156},
          doi = {10.3847/1538-4357/ad5a7e},
archivePrefix = {arXiv},
       eprint = {2402.13039},
 primaryClass = {astro-ph.SR},
       adsurl = {https://ui.adsabs.harvard.edu/abs/2024ApJ...970..156D}
}

@ARTICLE{Artymowicz_Lubow_1994,
       author = {{Artymowicz}, Pawel and {Lubow}, Stephen H.},
        title = "{Dynamics of Binary-Disk Interaction. I. Resonances and Disk Gap Sizes}",
      journal = {\apj},
         year = 1994,
        month = feb,
       volume = {421},
        pages = {651},
          doi = {10.1086/173679},
       adsurl = {https://ui.adsabs.harvard.edu/abs/1994ApJ...421..651A}
}

@ARTICLE{Armitage_Natarajan_2002,
       author = {{Armitage}, Philip J. and {Natarajan}, Priyamvada},
        title = "{Accretion during the Merger of Supermassive Black Holes}",
      journal = {\apjl},
         year = 2002,
        month = mar,
       volume = {567},
       number = {1},
        pages = {L9-L12},
          doi = {10.1086/339770},
archivePrefix = {arXiv},
       eprint = {astro-ph/0201318},
 primaryClass = {astro-ph},
       adsurl = {https://ui.adsabs.harvard.edu/abs/2002ApJ...567L...9A}
}

@ARTICLE{Munoz_Lai_2019,
       author = {{Mu{\~n}oz}, Diego J. and {Miranda}, Ryan and {Lai}, Dong},
        title = "{Hydrodynamics of Circumbinary Accretion: Angular Momentum Transfer and Binary Orbital Evolution}",
      journal = {\apj},
         year = 2019,
        month = jan,
       volume = {871},
       number = {1},
          eid = {84},
        pages = {84},
          doi = {10.3847/1538-4357/aaf867},
archivePrefix = {arXiv},
       eprint = {1810.04676},
 primaryClass = {astro-ph.HE},
       adsurl = {https://ui.adsabs.harvard.edu/abs/2019ApJ...871...84M}
}

@ARTICLE{Zrake_Tiede_MacF_Haiman_2021,
       author = {{Zrake}, Jonathan and {Tiede}, Christopher and {MacFadyen}, Andrew and {Haiman}, Zolt{\'a}n},
        title = "{Equilibrium Eccentricity of Accreting Binaries}",
      journal = {\apjl},
         year = 2021,
        month = mar,
       volume = {909},
       number = {1},
          eid = {L13},
        pages = {L13},
          doi = {10.3847/2041-8213/abdd1c},
archivePrefix = {arXiv},
       eprint = {2010.09707},
 primaryClass = {astro-ph.HE},
       adsurl = {https://ui.adsabs.harvard.edu/abs/2021ApJ...909L..13Z}
}

@ARTICLE{Shi+_2012,
       author = {{Shi}, Ji-Ming and {Krolik}, Julian H. and {Lubow}, Stephen H. and {Hawley}, John F.},
        title = "{Three-dimensional Magnetohydrodynamic Simulations of Circumbinary Accretion Disks: Disk Structures and Angular Momentum Transport}",
      journal = {\apj},
         year = 2012,
        month = apr,
       volume = {749},
       number = {2},
          eid = {118},
        pages = {118},
          doi = {10.1088/0004-637X/749/2/118},
archivePrefix = {arXiv},
       eprint = {1110.4866},
 primaryClass = {astro-ph.HE},
       adsurl = {https://ui.adsabs.harvard.edu/abs/2012ApJ...749..118S}
}

@ARTICLE{Siwek+_2023,
       author = {{Siwek}, Magdalena and {Weinberger}, Rainer and {Mu{\~n}oz}, Diego J. and {Hernquist}, Lars},
        title = "{Preferential accretion and circumbinary disc precession in eccentric binary systems}",
      journal = {\mnras},
         year = 2023,
        month = feb,
       volume = {518},
       number = {4},
        pages = {5059-5071},
          doi = {10.1093/mnras/stac3263},
archivePrefix = {arXiv},
       eprint = {2203.02514},
 primaryClass = {astro-ph.HE},
       adsurl = {https://ui.adsabs.harvard.edu/abs/2023MNRAS.518.5059S}
}

@article{Rafikov_2013,
	author = {Roman R. Rafikov},
	doi = {10.1088/0004-637X/774/2/144},
	journal = {The Astrophysical Journal},
	month = {aug},
	number = {2},
	pages = {144},
	publisher = {The American Astronomical Society},
	title = {STRUCTURE AND EVOLUTION OF CIRCUMBINARY DISKS AROUND SUPERMASSIVE BLACK HOLE BINARIES},
	url = {https://dx.doi.org/10.1088/0004-637X/774/2/144},
	volume = {774},
	year = {2013}}

@article{Rafikov_2016,
	author = {Roman R. Rafikov},
	doi = {10.3847/0004-637X/827/2/111},
	journal = {The Astrophysical Journal},
	month = {aug},
	number = {2},
	pages = {111},
	publisher = {The American Astronomical Society},
	title = {ACCRETION AND ORBITAL INSPIRAL IN GAS-ASSISTED SUPERMASSIVE BLACK HOLE BINARY MERGERS},
	url = {https://dx.doi.org/10.3847/0004-637X/827/2/111},
	volume = {827},
	year = {2016}}

@article{p1,
doi = {10.3847/1538-4357/ae29b0},
url = {https://doi.org/10.3847/1538-4357/ae29b0},
year = {2026},
month = {feb},
publisher = {The American Astronomical Society},
volume = {998},
number = {1},
pages = {4},
author = {Grcić, Marcela and D’Orazio, Daniel J. and Pessah, Martin E.},
title = {Insights from Analytical Theory of Eccentric Circumbinary Disks},
journal = {The Astrophysical Journal}
}

@ARTICLE{Noble+_2021,
       author = {{Noble}, Scott C. and {Krolik}, Julian H. and {Campanelli}, Manuela and {Zlochower}, Yosef and {Mundim}, Bruno C. and {Nakano}, Hiroyuki and {Zilh{\~a}o}, Miguel},
        title = "{Mass-ratio and Magnetic Flux Dependence of Modulated Accretion from Circumbinary Disks}",
      journal = {\apj},
         year = 2021,
        month = dec,
       volume = {922},
       number = {2},
          eid = {175},
        pages = {175},
          doi = {10.3847/1538-4357/ac2229},
archivePrefix = {arXiv},
       eprint = {2103.12100},
 primaryClass = {astro-ph.HE},
       adsurl = {https://ui.adsabs.harvard.edu/abs/2021ApJ...922..175N}
}

@ARTICLE{Lee_Dempsey_Lithwick_2019,
       author = {{Lee}, Wing-Kit and {Dempsey}, Adam M. and {Lithwick}, Yoram},
        title = "{Eccentric Modes in Disks with Pressure and Self-gravity}",
      journal = {\apj},
         year = 2019,
        month = feb,
       volume = {872},
       number = {2},
          eid = {184},
        pages = {184},
          doi = {10.3847/1538-4357/ab010c},
archivePrefix = {arXiv},
       eprint = {1811.11758},
 primaryClass = {astro-ph.EP},
       adsurl = {https://ui.adsabs.harvard.edu/abs/2019ApJ...872..184L}
}

@ARTICLE{Lee_Paele_2006,
       author = {{Lee}, Man Hoi and {Peale}, S.~J.},
        title = "{On the Orbits and Masses of the Satellites of the Pluto-Charon System}",
      journal = {arXiv e-prints},
         year = 2006,
        month = mar,
          eid = {astro-ph/0603214},
        pages = {astro-ph/0603214},
          doi = {10.48550/arXiv.astro-ph/0603214},
archivePrefix = {arXiv},
       eprint = {astro-ph/0603214},
 primaryClass = {astro-ph},
       adsurl = {https://ui.adsabs.harvard.edu/abs/2006astro.ph..3214L}
}

@ARTICLE{Dunhill+2015,
       author = {{Dunhill}, A.~C. and {Cuadra}, J. and {Dougados}, C.},
        title = "{Precession and accretion in circumbinary discs: the case of HD 104237}",
      journal = {\mnras},
         year = 2015,
        month = apr,
       volume = {448},
       number = {4},
        pages = {3545-3554},
          doi = {10.1093/mnras/stv28410.48550/arXiv.1411.0687},
archivePrefix = {arXiv},
       eprint = {1411.0687},
 primaryClass = {astro-ph.SR},
       adsurl = {https://ui.adsabs.harvard.edu/abs/2015MNRAS.448.3545D}
}

@ARTICLE{Lubow2022,
       author = {{Lubow}, Stephen H.},
        title = "{Forced eccentricity in circumbinary discs}",
      journal = {\mnras},
         year = 2022,
        month = nov,
       volume = {516},
       number = {4},
        pages = {5446-5453},
          doi = {10.1093/mnras/stac2636},
archivePrefix = {arXiv},
       eprint = {2209.06307},
 primaryClass = {astro-ph.SR},
       adsurl = {https://ui.adsabs.harvard.edu/abs/2022MNRAS.516.5446L}
}

@ARTICLE{ML2020,
       author = {{Mu{\~n}oz}, Diego J. and {Lithwick}, Yoram},
        title = "{Long-lived Eccentric Modes in Circumbinary Disks}",
      journal = {\apj},
         year = 2020,
        month = dec,
       volume = {905},
       number = {2},
          eid = {106},
        pages = {106},
          doi = {10.3847/1538-4357/abc74c},
archivePrefix = {arXiv},
       eprint = {2008.08085},
 primaryClass = {astro-ph.HE},
       adsurl = {https://ui.adsabs.harvard.edu/abs/2020ApJ...905..106M}
}

@ARTICLE{TO2016,
       author = {{Teyssandier}, Jean and {Ogilvie}, Gordon I.},
        title = "{Growth of eccentric modes in disc-planet interactions}",
      journal = {\mnras},
         year = 2016,
        month = may,
       volume = {458},
       number = {3},
        pages = {3221-3247},
          doi = {10.1093/mnras/stw521},
archivePrefix = {arXiv},
       eprint = {1603.00653},
 primaryClass = {astro-ph.EP},
       adsurl = {https://ui.adsabs.harvard.edu/abs/2016MNRAS.458.3221T}
}

@ARTICLE{GO2006,
       author = {{Goodchild}, Simon and {Ogilvie}, Gordon},
        title = "{The dynamics of eccentric accretion discs in superhump systems}",
      journal = {\mnras},
         year = 2006,
        month = may,
       volume = {368},
       number = {3},
        pages = {1123-1131},
          doi = {10.1111/j.1365-2966.2006.10197.x},
archivePrefix = {arXiv},
       eprint = {astro-ph/0602492},
 primaryClass = {astro-ph},
       adsurl = {https://ui.adsabs.harvard.edu/abs/2006MNRAS.368.1123G}
}

@ARTICLE{Dan-gaps,
       author = {{D'Orazio}, Daniel J. and {Haiman}, Zolt{\'a}n and {Duffell}, Paul and {MacFadyen}, Andrew and {Farris}, Brian},
        title = "{A transition in circumbinary accretion discs at a binary mass ratio of 1:25}",
      journal = {\mnras},
         year = 2016,
        month = jul,
       volume = {459},
       number = {3},
        pages = {2379-2393},
          doi = {10.1093/mnras/stw792},
archivePrefix = {arXiv},
       eprint = {1512.05788},
 primaryClass = {astro-ph.HE},
       adsurl = {https://ui.adsabs.harvard.edu/abs/2016MNRAS.459.2379D}
}

@ARTICLE{Dan-transition,
       author = {{D'Orazio}, Daniel J. and {Duffell}, Paul C.},
        title = "{Orbital Evolution of Equal-mass Eccentric Binaries due to a Gas Disk: Eccentric Inspirals and Circular Outspirals}",
      journal = {\apjl},
         year = 2021,
        month = jun,
       volume = {914},
       number = {1},
          eid = {L21},
        pages = {L21},
          doi = {10.3847/2041-8213/ac0621},
archivePrefix = {arXiv},
       eprint = {2103.09251},
 primaryClass = {astro-ph.HE},
       adsurl = {https://ui.adsabs.harvard.edu/abs/2021ApJ...914L..21D}
}

@article{Ogilvie_2001,
	doi = {10.1046/j.1365-8711.2001.04416.x},
	url = {https://doi.org/10.1046%2Fj.1365-8711.2001.04416.x},
	year = 2001,
	month = {jul},
	publisher = {Oxford University Press ({OUP})},
	volume = {325},
	number = {1},
	pages = {231--248},
	author = {G. I. Ogilvie},
	title = {Non-linear fluid dynamics of eccentric discs},
	journal = {Monthly Notices of the Royal Astronomical Society}
}

@ARTICLE{2008Planetesimal,
       author = {{Paardekooper}, S. -J. and {Th{\'e}bault}, P. and {Mellema}, G.},
        title = "{Planetesimal and gas dynamics in binaries}",
      journal = {\mnras},
         year = 2008,
        month = may,
       volume = {386},
       number = {2},
        pages = {973-988},
          doi = {10.1111/j.1365-2966.2008.13080.x},
archivePrefix = {arXiv},
       eprint = {0802.0927},
 primaryClass = {astro-ph},
       adsurl = {https://ui.adsabs.harvard.edu/abs/2008MNRAS.386..973P}
}

@ARTICLE{Miranda_Munoz_Lai_2017,
       author = {{Miranda}, Ryan and {Mu{\~n}oz}, Diego J. and {Lai}, Dong},
        title = "{Viscous hydrodynamics simulations of circumbinary accretion discs: variability, quasi-steady state and angular momentum transfer}",
      journal = {\mnras},
         year = 2017,
        month = apr,
       volume = {466},
       number = {1},
        pages = {1170-1191},
          doi = {10.1093/mnras/stw3189},
archivePrefix = {arXiv},
       eprint = {1610.07263},
 primaryClass = {astro-ph.SR},
       adsurl = {https://ui.adsabs.harvard.edu/abs/2017MNRAS.466.1170M}
}

@ARTICLE{Bromley_Kenyon_2015,
       author = {{Bromley}, Benjamin C. and {Kenyon}, Scott J.},
        title = "{Planet Formation around Binary Stars: Tatooine Made Easy}",
      journal = {\apj},
         year = 2015,
        month = jun,
       volume = {806},
       number = {1},
          eid = {98},
        pages = {98},
          doi = {10.1088/0004-637X/806/1/98},
archivePrefix = {arXiv},
       eprint = {1503.03876},
 primaryClass = {astro-ph.EP},
       adsurl = {https://ui.adsabs.harvard.edu/abs/2015ApJ...806...98B}
}

@ARTICLE{MacFadyen_Milosavljevic_2008,
       author = {{MacFadyen}, Andrew I. and {Milosavljevi{\'c}}, Milo{\v{s}}},
        title = "{An Eccentric Circumbinary Accretion Disk and the Detection of Binary Massive Black Holes}",
      journal = {\apj},
         year = 2008,
        month = jan,
       volume = {672},
       number = {1},
        pages = {83-93},
          doi = {10.1086/523869},
archivePrefix = {arXiv},
       eprint = {astro-ph/0607467},
 primaryClass = {astro-ph},
       adsurl = {https://ui.adsabs.harvard.edu/abs/2008ApJ...672...83M}
}

@ARTICLE{Dittmann_Ryan_2024,
       author = {{Dittmann}, Alexander J. and {Ryan}, Geoffrey},
        title = "{The Evolution of Accreting Binaries: From Brown Dwarfs to Supermassive Black Holes}",
      journal = {\apj},
         year = 2024,
        month = may,
       volume = {967},
       number = {1},
          eid = {12},
        pages = {12},
          doi = {10.3847/1538-4357/ad2f1e},
archivePrefix = {arXiv},
       eprint = {2310.07758},
 primaryClass = {astro-ph.GA},
       adsurl = {https://ui.adsabs.harvard.edu/abs/2024ApJ...967...12D}
}

@ARTICLE{Thun_plus_2017,
       author = {{Thun}, Daniel and {Kley}, Wilhelm and {Picogna}, Giovanni},
        title = "{Circumbinary discs: Numerical and physical behaviour}",
      journal = {\aap},
         year = 2017,
        month = aug,
       volume = {604},
          eid = {A102},
        pages = {A102},
          doi = {10.1051/0004-6361/201730666},
archivePrefix = {arXiv},
       eprint = {1704.08130},
 primaryClass = {astro-ph.EP},
       adsurl = {https://ui.adsabs.harvard.edu/abs/2017A&A...604A.102T}
}

@ARTICLE{Farris_Duffel_2014,
       author = {{Farris}, Brian D. and {Duffell}, Paul and {MacFadyen}, Andrew I. and {Haiman}, Zoltan},
        title = "{Binary Black Hole Accretion from a Circumbinary Disk: Gas Dynamics inside the Central Cavity}",
      journal = {\apj},
         year = 2014,
        month = mar,
       volume = {783},
       number = {2},
          eid = {134},
        pages = {134},
          doi = {10.1088/0004-637X/783/2/134},
archivePrefix = {arXiv},
       eprint = {1310.0492},
 primaryClass = {astro-ph.HE},
       adsurl = {https://ui.adsabs.harvard.edu/abs/2014ApJ...783..134F}
}

@ARTICLE{Dorazio_Haiman_2013,
       author = {{D'Orazio}, Daniel J. and {Haiman}, Zolt{\'a}n and {MacFadyen}, Andrew},
        title = "{Accretion into the central cavity of a circumbinary disc}",
      journal = {\mnras},
         year = 2013,
        month = dec,
       volume = {436},
       number = {4},
        pages = {2997-3020},
          doi = {10.1093/mnras/stt1787},
archivePrefix = {arXiv},
       eprint = {1210.0536},
 primaryClass = {astro-ph.GA},
       adsurl = {https://ui.adsabs.harvard.edu/abs/2013MNRAS.436.2997D}
}

@ARTICLE{Miranda_Rafikov_2018,
       author = {{Miranda}, Ryan and {Rafikov}, Roman R.},
        title = "{Fast and Slow Precession of Gaseous Debris Disks around Planet-accreting White Dwarfs}",
      journal = {\apj},
         year = 2018,
        month = apr,
       volume = {857},
       number = {2},
          eid = {135},
        pages = {135},
          doi = {10.3847/1538-4357/aab9a2},
archivePrefix = {arXiv},
       eprint = {1802.00464},
 primaryClass = {astro-ph.EP},
       adsurl = {https://ui.adsabs.harvard.edu/abs/2018ApJ...857..135M}
}
\appendix{}
\label{app1}

\section{The Gravitational potential of a disk}\label{appendix a}

We assume that the mass of the disk is high enough to make $\Phi_\mathrm{d}$, the gravitational potential caused by the disk mass, a non-negligible contribution to the total gravitational potential. The total gravitational potential is then:
\begin{equation}
    \Phi=\Phi_\mathrm{b}+\Phi_\mathrm{d},
\end{equation}
$\Phi_\mathrm{b}$ is the potential from the binary given by Eq. (\ref{eq: potential binary m=0}). We assume that the only relevant gravitational potential from the disk is its $m=0$ component. 

The disk considered here contains a central cavity. Consequently, the disk’s gravitational potential has different forms in the disk body and within the cavity.
The effects of that potential on the system, too, are different in the two regions. Inside the disk, including the gravitational potential caused by the disk mass changes the free precession of the disk. In the cavity, the gravitational potential caused by the disk mass causes the precession of the binary. For this reason, we treat the two cases separately.

\subsection{The Gravitational Potential Inside the Disk and Free CBD Modes}

 The potential inside the disk, for $\rin<r<\rout$, is different for different values of the power-law-density exponent $\zeta$. For $\zeta=-1$, it is \citep{Hure_Hersant_2008}:
\begin{equation}\label{eq: potential for power law finite disk, zeta=-1}
    \Phi_\mathrm{d}(r)\approx 2\pi G\Sigma(\rout) \rout \Bigg[ \ln{\left(\frac{r}{\rout}\right)}+\frac{\rin}{r}-1.4\Bigg],
\end{equation}
 for $\zeta=-2$, it is:
\begin{equation}\label{eq: potential for power law finite disk, zeta=-2}
    \Phi_\mathrm{d}(r)\approx 2\pi G\Sigma(\rout) \rout \Bigg[\ln{\left(\frac{\rin}{r}\right)}+\frac{r}{\rout}-1.4\Bigg]\frac{\rout}{r}.
\end{equation}
For all other values of $\zeta$ the potential is given by:
\begin{equation}\label{eq: potential for power law finite disk}
    \begin{split}
        \Phi_\mathrm{d}\approx&-2\pi G\Sigma(\rout) \rout\Bigg[ \frac{1}{1+\zeta}
        + \Big[0.4-\frac{1}{(1+\zeta)(2+\zeta)}\Big] \left( \frac{r}{\rout}\right)^{1+\zeta} 
        -\frac{1}{2+\zeta}\left( \frac{\rin}{\rout}\right)^{1+\zeta}\frac{\rin}{r}
        \Bigg]
    \end{split},
\end{equation}
where we have retained only the three lowest order terms in the series expansion of the potential in radius. \cite{Hure_Hersant_2008} provide additional terms in potential expansion, and show that the expressions given by Eq. (\ref{eq: potential for power law finite disk, zeta=-1})-(\ref{eq: potential for power law finite disk}) differ from exact numerical solutions by at most a few percent. Additional analytical and numerical results for potentials of infinite and finite power-law-density disks can be found in \cite{Hure_Pierens_2005}, \cite{Hure_Pelat_2007}, and \cite{Hure_Hersant_2007}.

We substitute Eqs. (\ref{eq: potential for power law finite disk, zeta=-1})-(\ref{eq: potential for power law finite disk}) into Eq. (\ref{eq: gravitational forcing term for the eccentricty equation}) to find the non-Keplerian gravitational CBD eccentricity contribution from the disk:
\begin{equation}\label{eq: f_gd}
    f_{\mathrm{g},m=0}=i\Sigma(r) E \pi \left[0.4(1+\zeta)(2+\zeta)-1  \right] \ob^2 \ab^2 \left(\frac{r}{\ab}\right)^{\zeta+1}.
\end{equation}
Equation (\ref{eq: f_gd}) is valid for all values of $\zeta$ for $\rin<r<\rout$.

\subsection{The Gravitational Potential Inside the Disk Cavity}

The binary is located within the disk cavity, at $r< \rin$, for which the gravitational potential from the power-law mass distribution in the disk is \citep{Hure_Hersant_2008}:
\begin{equation}\label{eq: grav potential inside cavity , power-law disk}
\begin{split}
    \Phi_\mathrm{d}(r)\approx& 2\pi G \Sigma(\rout)\rout \sum_{n=0}^\infty a_n \left(\frac{r}{\rout}\right)^{2n},
\end{split}
\end{equation}
where the coefficients $a_n$ are given as \citep{Hure_Hersant_2008}:
\begin{equation}
a_n=\frac{\gamma_n}{2n-\zeta-1}\left[1-\left(\frac{\rin}{\rout}\right)^{1+\zeta-2n} \right],
\end{equation}
with coefficients $\gamma_n$ given as a recursion $\gamma_n=\gamma_{n-1}(2n-1)^2/(2n)^2$ with $\gamma_0=1$.

We assume that the gravitational potential of the massive disk exerts a force on the binary, and we neglect non-axisymmetric contributions, considering only the radial component of the force.
If a small radial forcing $f_r$ poses a perturbation to two-body-problem binary evolution, the binary precession rate is given as \citep{Murray_Dermott_1999}:
\begin{equation}\label{eq: bin precession equation in terms of f_r}
    \wb=\frac{\sqrt{\ab (1-\eb^2)}}{\eb \sqrt{G\Mb}} \left[-f_r\cos f \right],
\end{equation}
where the external radial forcing on the binary $f_r$ is caused by the gravitational potential of the disk $\Phi_\mathrm{d}$:
\begin{equation} \label{eq: radial force of a massive disk on binary}
    f_r=-\frac{\partial \Phi_\mathrm{d}}{\partial r}.
\end{equation}

Substituting Eq. (\ref{eq: grav potential inside cavity , power-law disk}) into Eq. (\ref{eq: radial force of a massive disk on binary}), we obtain:
\begin{equation}\label{eq: f_r for power-law disks}
    f_r=-2 \pi G \Sigma(\rout) \left(2a_1\frac{r}{\rout}+4a_2\frac{r^3}{\rout^3} \right) .
\end{equation}

Substituting Eq. (\ref{eq: f_r for power-law disks}) into Eq. (\ref{eq: bin precession equation in terms of f_r}), we obtain the precession rate of the binary:
\begin{equation}\label{eq: binary precession rate}
\begin{split}
     \wb=&\frac{\sqrt{ 1-\eb^2}}{\eb} 2 \pi \ob \qd \left(\frac{\rout}{\ab}\right)^\zeta \left(2a_1\frac{r}{\rout}+4a_2\frac{r^3}{\rout^3} \right) \cos{f} ,
\end{split}
\end{equation}
where we have used $\ob=\sqrt{G\Mb/\ab^{3}}$ and Eq. (\ref{eq: q_d}).

We use the series expansion for radius
in powers of the binary eccentricity in terms of the mean anomaly $M$ given by (see \cite{Murray_Dermott_1999})
\begin{equation}
    \begin{split}
        \frac{r}{\ab}=&  1-\eb \cos{M}+\eb^2 \frac{1-\cos{2M}}{2}+ \eb^3\frac{3}{8}\left[\cos{M}-\cos{3M} \right]      \\
    \end{split}
\end{equation}
and the true anomaly: 
\begin{equation}
\begin{split}
    \cos{f}=&\cos{M}+\eb \left[\cos{2M}-1 \right]+\eb^2 \frac{9}{8}\left[\cos{3M}-\cos{M} \right]+\eb^3\frac{4}{3}\left[\cos{4M}-\cos{2M} \right],
    \end{split}
\end{equation}
to average out Eq. (\ref{eq: binary precession rate}) over the mean anomaly:
\begin{equation}\label{eq: average binary precession rate}
\begin{split}
    \frac{\bomb}{\ob}=&-2\pi \qd \sqrt{1-\eb^2}\Bigg[3a_1 \left(\frac{\rout}{\ab} \right)^{\zeta-1}+a_2\left( \frac{\rout}{\ab}\right)^{\zeta-3}\left(10+\frac{15}{2}\eb^2\right) \Bigg],
    \end{split}
\end{equation}
where 
\begin{equation}
\bomb\equiv \frac{1}{2\pi}\int_0^{2\pi}\wb \, dM.
\end{equation}

\subsection{The Comparison to Free CBD Eccentric Modes}\label{subsec: free modes}

We briefly look into the effect of the disk mass on the free eccentric CBD modes, and compare it to the forced eccentric CBD modes. To find the free modes, we neglect the forcing term in Eq. (\ref{eq: gravitational forcing term w_bin}), and include the disk mass in the eccentricity equation through the modified $m=0$ gravitational potential in the disk only. We solve the resulting unforced eccentricity equation numerically for a 2D locally isothermal disk. 

In Figure \ref{fig: w_0(q_d) free}, we plot solutions for the ground mode precession frequency of a free mode. Unlike the forced modes (Eq. (\ref{eq: average binary precession })), the precession frequency of free modes decreases with increasing the disk-to-binary mass ratio.

To understand this, similarly to $\wq$, we define $\omega_\mathrm{d}$ as the precession frequency of the disk if pressure and binary gravitational potentials are neglected; we set $p=\Phi_\mathrm{Q}=\Phi_{m=1}=0$ into Eqs. (\ref{eq: eccentricity equation in terms of f_p and f_g with forcing})-(\ref{eq: gravitational forcing term w_bin})):
\begin{equation}\label{eq: w_d - zeta}
    \omega_\mathrm{d}(r)=\pi \left[ 0.4(1+\zeta)(2+\zeta)-1\right]  \times \qd \left(\frac{r}{\ab}\right)^{\zeta+1/2} \ob.
\end{equation}
For power-law-density exponents used in this paper ($\zeta=-1/2,-1,-3/2$), the precession frequency $\omega_\mathrm{d}(r)$ induced by the disk mass is negative. For the free mode precession frequency, this is reflected as the mode frequency decreasing with increased disk mass. For $\zeta=-1/2$, $\omega_\mathrm{d}$ is a constant, so we can find $\omega(\qd, \zeta=-1/2)\approx \omega(\qd=0,\zeta=-1/2)-2.2 \qd \ob$.

\begin{figure}\centering
    \includegraphics[scale=0.97]{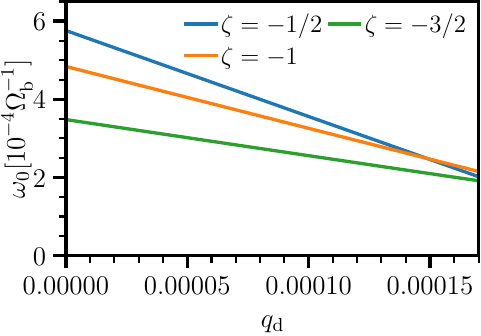}  
    \caption{The precession frequency of a free ground ($n$=0) CBD mode $\omega_0$ for three values of the power-law-density exponent ($\zeta=-1/2$ (solid blue line), $\zeta=-1$ (solid orange line), and $\zeta=-3/2$ (solid green line)).  The disk is a 2D locally isothermal disk with inner and outer disk radii $\rin=2.5\ab$ and $\rout=200\ab$. The binary eccentricity and mass ratio are $\eb=0.07$ and $\qb=0.6$, and the disk apsect ratio is $h=0.1$. The free mode precession frequency decreases with increasing disk-to-binary mass ratio.}\label{fig: w_0(q_d) free}
\end{figure}

\end{document}